\pdfoutput=1
\documentclass[manuscript]{aastex6}

\usepackage{natbib}
\usepackage{color}
\usepackage{graphicx}
\usepackage{rotating}
\usepackage{epstopdf}
\usepackage{gensymb}
\usepackage{epsfig}
\usepackage{subfigure}

\begin{document}
\title{Automated Detection of Coronal Mass Ejections in STEREO Heliospheric Imager  data}
\author{V.~Pant$^{1}$,
        S. Willems$^{2}$,
        L. Rodriguez$^{2}$
        M. Mierla$^{2,3}$
        D. Banerjee$^{1,4}$
        J. A. Davies$^{5}$
        }
\affil{$^{1}$ Indian Institute of Astrophysics, Bangalore-560 034, India\\
         $^{2}$Solar-Terrestrial Center of Excellence, Royal Observatory of Belgium, Avenue Circulaire 3, B-1180 Brussels, Belgium\\
          $^{3}$Institute of Geodynamics of the Romanian Academy, Bucharest, Romania\\
           $^{4}$Center of Excellence in Space Sciences, IISER Kolkata, India \\
           $^{5}$RAL Space, STFC Rutherford Appleton Laboratory, Harwell Campus, OX11 OQX, UK}

\begin{abstract}
We have performed, for the first time, the successful automated detection of Coronal Mass Ejections (CMEs) in data from the inner heliospheric imager (HI-1) cameras on the STEREO A spacecraft. Detection of CMEs is done in time-height maps based on the application of the Hough transform, using a modified version of the CACTus software package, conventionally applied to coronagraph data. In this paper we describe the method of detection. We present the result of the application of the technique to a few CMEs that are well detected in the HI-1 imagery, and compare these results with those based on manual cataloging methodologies. We discuss in detail the advantages and disadvantages of this method. 
\end{abstract}
\newpage



\section{Introduction}
According to the original definition, a Coronal Mass Ejection (CMEs) is an observable changes in coronal structure that occurs on time scales between a few minutes to several hours and involves the appearance \citep{1984JGR....89.2639H} and outward propagation \citep{1996Ap&SS.243..187S} of new, discrete and bright white-light features in the coronagraph field of view. CMEs result from the episodic expulsion of plasma and magnetic field from the solar atmosphere into the heliosphere with speeds that are typically 400 km~s$^{-1}$ but that can range from 100 to 2500~km~s$^{-1}$ \citep{2004JGRA..109.7105Y,2011JASTP..73..671M}. CMEs are considered the most energetic events in the solar system. Furthermore they are very important in terms of space weather, being the drivers of the largest geomagnetic storms detected at Earth. Since 1996, we have been able to monitor CMEs routinely from the L1 vantage point using the Large Angle Spectrometric Coronagraph (LASCO) \citep{1995SoPh..162..357B} on the Solar and Heliospheric Observatory (SOHO) spacecraft. Moreover, since late 2006,  we have also been able to monitor CMEs from a location off the Sun-Earth line using the COR1 and COR2 coronagraphs that form part of the Sun Earth Connection Coronal and Heliospheric Investigation (SECCHI) \citep{2008SSRv..136...67H} imaging package on the twin-spacecraft Solar Terrestrial Relation Observatory (STEREO) mission. The SECCHI Heliospheric Imager (HI) instruments on STEREO effectively extend the coronagraph observing methodology out to larger distances from the Sun by providing wide-angle white-light imaging of the heliosphere out to 1 AU and beyond \citep{2009SoPh..254..387E, 2009SoPh..256..219H}. The HI instrument on each STEREO spacecraft comprises of two camera, HI-1 and HI-2. The angular fields of view (FOVs) of HI-1 and HI-2 are 20$\degree$ and 70$\degree$, with the FOVs being centered, in nominal operations, at around 14$\degree$ and 54$\degree$ elongation, respectively \citep{2000SPIE.4139..284S}. The extensive HI FOV allows us to observe CMEs propagating over vast distances of interplanetary space \citep{2007SPIE.6689E..07E,2009GeoRL..36.8102D,2008SoPh..247..171H,2009SoPh..256..219H}. The concept of wide-angle heliospheric imaging was first demonstrated by the Solar Mass Ejection Imager (SMEI) onboard the Earth-orbiting  {\it Coriolis} spacecraft \citep{2003SoPh..217..319E}. The terminology of interplanetary CMEs (ICMEs) is often applied to the interplanetary counterparts of CMEs \citep[see,][]{2000GeoRL..27..145G,2006ApJ...647..648R,2006SSRv..123...31Z}. However, due to the success of {\it Coriolis}/SMEI and STEREO/HI in filling the vast observational gap between coronagraph imagery and in situ measurements, some authors are now suggesting that the terminology CME should be applied to both phenomena \citep{2012LRSP....9....3W,2013SoPh..285....1B}. \\
Since the launch of STEREO, CMEs have been identified through visual inspection of HI images and the resultant event catalogs have been made public \citep{2014SpWea..12..657B,2015SpWea..13..709B,harrison2016}. However, such visual identification of CMEs is biased by human subjectivity and hence CME detection may or may not be consistent over an extended period of time \citep{2014ApJ...784L..27W}. It is worth mentioning that each CME in the Solar Stormwatch CME catalog \citet{2014SpWea..12..657B} is identified manually by multiple independent operators in order to reduce the subjectivity of the detection. The properties of a given CME  are derived by averaging the independent detections. Nevertheless, automated detection offers the capability of providing more objective CME detection. One such software package, Computer Aided CME Tracking (CACTus) was developed to detect CMEs in LASCO/C2 and C3 coronagraph images  \citep{2004A&A...425.1097R} and subsequently applied to COR2 coronagraph imagers from  STEREO. CACTus applies the original definition of a CME; ``a CME is a new, discrete,  bright, white-light  feature in  the  coronagraph FOV with a radially outward velocity" \citep{2004A&A...425.1097R}. CME detection using CACTus is more objective and faster compared with the visual identification because CMEs are detected, and characterised according to a strict set of precisely defined constraints. The catalogs produced using CACTus (available online \footnote{\url{http://sidc.oma.be/cactus}}) are similar to manually compiled catalogs in terms of parameters that they contain. \citet{2009ApJ...691.1222R} have compared the CME parameters derived by CACTus with those obtained by manual detection of CMEs in LASCO/C2 and C3 images. Apart from CACTus, there are several other catalogs of CMEs that have been identified automatically in coronagraph data. The Coronal Image Processing (CORIMP) automated detection algorithm uses normalized radial gradient filtering and deconvolution to separate quiescent structures (background corona) and dynamic structures (features such as CMEs, that propagate radially outward) \citep{2012ApJ...752..144M,2012ApJ...752..145B}. CORIMP has been used to detect CMEs automatically in LASCO and COR2 images. Another such algorithm, Automatic Recognition of Transient Events and Marseille Inventory from Synoptic maps (ARTEMIS), was developed by \citet{2009SoPh..257..125B}. ARTEMIS detects CMEs automatically in LASCO synoptic maps using image filtering and segmentation techniques. Furthermore, the Solar Eruptive Events Detection System (SEEDS) also detects CMEs in polar transformed running difference images from LASCO and COR2  \citep{2008SoPh..248..485O}. SEEDS isolates the leading edges of CMEs by intensity thresholding; by tracking CMEs in sequential images, their speeds and accelerations are calculated. \citet{2015JSWSC...5A..19B}  compared the CORIMP catalogs with catalogs generated using other automated ({\it i.e,} CACTus and SEEDS) and manual detection methods ({\it i.e} the CDAW catalog). The authors demonstrated the robustness of CORIMP in deriving the kinematics of the automatically-detected CMEs. The automated detection of CMEs in the heliosphere is, however, not an easy task, mainly due to their low brightness compared to that of the other contributions to the white-light signal (principally the F-coronal and stellar backgrounds). Despite this, there have been previous attempts to automatically identify CMEs in heliospheric images. The Automated Interplanetary Coronal Mass Ejection Detection (AICMED) tool was developed by \citet{2012JGRA..117.5103T} to detect CMEs in SMEI data, in particular. Like CACTus, AICMED works on the principle of the Hough transform, which detects straight ridges. AICMED uses the Hough transform to detect curved ridges in time-elongation maps (commonly called J-maps) by splitting each curved ridge into several straight ridges (this curvature, as will be discussed later, is a geometric artefact associated with imaging out to large elongations). However due to a number of features in the SMEI data (such as rings generated by hot pixels, scattered light from the moon, cometary tails, high-altitude aurora and particle hits during crossings of the South Atlantic Anomaly (SAA) and auroral zones) there were many false detections. Recently \citet{2015SpWea..13..709B} reported the differences in velocities estimated using automated and manual tracking of ridges in time-elongation maps created from STEREO/HI-1 and HI-2 images \citep[see also,][]{2009AnGeo..27.4349S}, although the authors did not perform a completely automatic detection of CMEs in the HI imagery. \\
We have adapted the CACTus software package to automatically detect CMEs in STEREO/HI-1 data. In this manuscript we first describe the methodology that we have used (section~\ref{method}). Subsequently in sections~\ref{mancatalog} and \ref{sec4}, we compare the automatically-derived parameters for a selection of CMEs with analogous entries in a manual catalogs for a selection of CMEs: namely the time of appearance (t$_{0}$), central position angle (CPA) of propagation, position angle width (da) and the projected speed (v).
\section{Method of Detection}
\label{method}
In this section we describe the method of automated detection of CMEs in HI-1 data from STEREO-A, which is an adaptation of the original CACTus methodology as discussed in \citet{2004A&A...425.1097R} but with some modifications in order to make it work with the HI images. CME detection by CACTus is based on the principle of the Hough transform \citep[see,][]{1997ipsa.book.....J}, which can be used to detect straight lines in noisy data. The brightness of CMEs in heliospheric images is generally lower relative to other contributions to the signal (such as F-coronal and stellar backgrounds) than in coronagraph images, leading to an inferior signal to noise ratio. Since the automated detection of CMEs depends critically in their clarity, we need to carefully process the images before implementing the Hough transform to the time-height maps.
\subsection{Pre-processing of HI Images}
\begin{figure*}[h]
\centering
\includegraphics[scale=0.55]{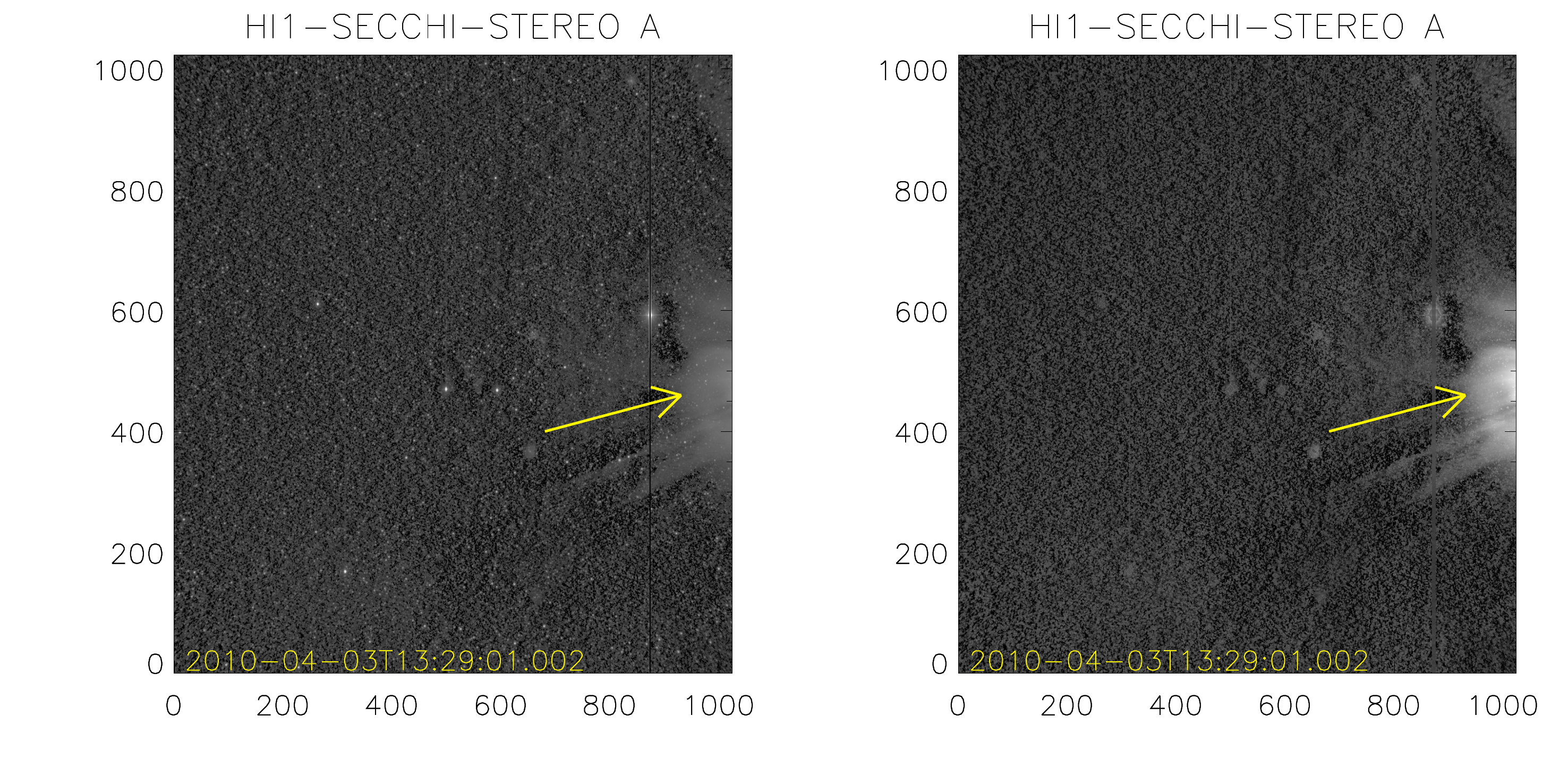}
\caption{\textrm{Left}: Level 2 STEREO-A HI-1 image after 1-day background subtraction and the initial removal of bright streaks (most clearly that resulting from the presence of Mercury in the FOV). \textrm{Right}: As \textrm{left}, but with further processing to reduce the effects of bright planets and stars (as discussed in text). The yellow arrow indicates the leading edge of a CME (at a PA of 102$\degree$) observed on 2010/04/03.}
\label{fig1} 
\end{figure*}

\begin{itemize}
\item \textbf{Data acquisition:} The 1-day background-subtracted level 2 HI-1 science images (array size, 1024 $\times$ 1024) in units of DN sec$^{-1} $ per CCD pixel are downloaded via the UK Solar System Data Center (UKSSDC) website \footnote{ \url{http://www.ukssdc.rl.ac.uk/solar/stereo/data.html}} (note that pixels on the CCD detector are binned 2~$\times$~2 onboard to generate the science images; our subsequent use of the term pixel refers to a pixel in a science image). These images have a nominal cadence of 40 minutes. For illustration throughout section~\ref{method}, we analyse HI-1 images from the period extending from the beginning of 2010/04/02 to the end of 2010/04/04, encompassing about 72 hours. However, in section~\ref{mancatalog} when comparing with the results of manual cataloging, we include five additional days in our analysis. The background that is subtracted from each HI-1 image is the average of the lowest quartile of the data in each pixel within the FOV over a 1-day period centered on the image of interest \footnote{\url{http://www.ukssdc.rl.ac.uk/solar/stereo/documentation/HI_processing_L2_data.html}} \citep{2016SoPh}. By subtracting a daily background, quasi-static components of the signal, principally the F-corona and also the less variable elements of the K-corona (eg, streamers) are removed. Such a 1-day background subtracted level 2 HI-1 image from STEREO-A is shown in the left hand panel of figure~\ref{fig1}. The processing that has been applied to the images is described on the UKSSDC website \footnote{ \url{http://www.ukssdc.rl.ac.uk/solar/stereo/documentation/HI_processing.html}}.
\item \textbf{Removal of stars and planets:} It is evident from this image that the white-light signal due to the presence of bright stars and planets in the FOV can exceed the CME signal (the leading edge of the CME is marked with a yellow arrow). The stars and planets are not removed by the background-subtraction procedure as they move through the FOV at a relatively fast rate. To reduce the effects of the stars and the planets we use a sigma filter, which works by first computing the mean and the standard deviation of the intensity of the neighbouring pixels excluding the pixel on which it is centered. If the intensity of the central pixel is greater than a chosen threshold (mean + 4~$\times$~standard deviation), its value is replaced by the mean value of its neighbouring pixels. This process is iterated over all of the pixels in an image, recursively upto 20 times or until no further change is observed. By examining few test cases, we determined that using a sigma filtering with a width of 50 pixels works well at reducing the effects of planets and stars.
\item \textbf{Removal of bright streaks associated with planets and bright stars:} Unlike for coronagraphs with their smaller FOVs, in which only a few planets and bright stars tend to be present at one time, more planets and bright stars are present in the somewhat larger HI-1 FOV. Planets and bright stars are associated with bright vertically-extended streaks, which result from vertical blooming of the signal in saturated pixels. It is crucial to remove these bright vertical streaks prior to CME detection. It should be noted that the {\it secchi\_prep.pro} routine (which is applied to the HI data as part of the processing performed at the UKSSDC) attempts to replace the bright vertically extended streaks with NaN values (NaN streak). However, we find that this procedure is often not always totally effective. An incompletely-removed bright streak may still be brighter than a CME in the difference images that are used for CME detection by CACTus (see section~\ref{polarsect}) and therefore may affect the CME detection. Hence, CACTus performs additional processing of the HI-1 data, which is not present in earlier versions of the software in order to remove these residual bright streaks. This process involves the use of dilation. Dilation is an image processing technique that ``grows" or ``widens" an object. The extent of the widening is based on the shape of the kernel \citep{2002dip..book.....G}. We choose a one dimensional kernel with a horizontal width of 6 pixels for dilation. This width of kernel is found to work well at covering any residual bright vertical streaks associated with a planet or bright star. Upon its application the original NaN streak that is implemented in the UKSSDC processing becomes wider, as a result of which any residual bright streak is completely replaced by NaNs. This process results in the highly effective replacement of the residual bright streaks in the HI-1 images. Subsequently, we use convolution based on a one dimensional kernel of 100 pixels in width to replace the NaN streak with values of the neighbouring (horizontal) pixels (note that the NaN streak extends over the entire vertical range of the image). Through this procedure, we isolate the position of the dilated NaN streak and replace it with the intensity obtained through convolution. This fills the affected pixels with the values of surrounding pixels without smoothing the entire image. 
\end{itemize}
The processed image after the application of the above procedures is shown in right hand panel of figure~\ref{fig1}. The CME is now brighter relative to the stars and planets which are suppressed and the residual bright streak (with its associated NaN streak) is now filled with neighbouring intensity values.

\subsection{Polar Transformation}
\label{polarsect}
The next step towards the detection of CMEs in STEREO/HI imagery is the conversion of the helioprojective cartesian (HPC) coordinates provided as one of the standard coordinate system of the HI images, into polar ($\theta-r$) coordinates, where $\theta$ is the Position Angle (PA), measured counterclockwise from Solar north, and $r$ is the distance from Sun center projected onto the plane of sky (POS) as seen from the spacecraft.
\begin{itemize}
\item In order to do this, we first convert the HPC coordinates of the HI images to helioprojective-radial (HPR) coordinates \citep[see][]{2006A&A...449..791T} using routines that are available in the SolarSoft package. It should be noted that the accurate pointing calibration of the HI imagery is achieved in the UKSSDC preprocessing through comparison of the star field within each image with a star catalog \citep{2009SoPh..254..185B}. 
\item The conversion to HPR, assigns each pixel in the HI image with two values, the PA ($\theta$) and the elongation ($\epsilon$), the angle between its line-of-sight and the spacecraft-Sun line. Knowing the elongation of each pixel in the image, we can calculate its projected distance from Sun-centre on the POS, using the expression
\begin{equation}
\label{eq1}
r=d~\tan(\epsilon),
\end{equation}
where $d$ is the distance of the observing spacecraft from the Sun. 
\item The polar ($\theta-r$) representation of the image presented in the right hand panel of figure~\ref{fig1} is shown in left hand panel of figure~\ref{fig2}, where the x axis represents PA ($\theta$) and the y axis represents the projected distance in the POS derived using Equation~\ref{eq1}. The data are binned such that one height bin corresponds to 100000 km (roughly 1/7$^{th}$ of solar radius). Since CMEs are large scale features, this binning does not lose any structural information. Furthermore, binning reduces noise and computation time.
\item To remove quasi-static features that vary on a time-scale shorter than the 1-day subtracted background, and longer than the 40 minute image cadence of HI-1, running difference $\theta-r$ images are generated automatically in a manner analogous to what is done in the conventional version of CACTus \citet{2004A&A...425.1097R}. In particular, this removes much of the shorter time-scale variation of the streamer belt. The right hand panel of figure~\ref{fig2} shows only the positive values of the difference image (corresponding to an increase in brightness, hence density, relative to the previous image). 
\end{itemize}

\subsubsection{Geometric models and fitting methods}
\label{gm}
Since the advent of wide-angle imaging of the solar wind, by {\it Coriolis}/SMEI and STEREO/HI, a number of geometries have been derived that enable the elongation angle of a CME (or other solar wind structure) to be converted to a radial distance from Sun-centre, namely Point-P (PP) \citep{2006JGRA..111.4105H}, Fixed-$\phi$ (FP) \citep{2007JGRA..112.9103K, 2008GeoRL..3510110R, 2008ApJ...675..853S}, Harmonic Mean (HM) \citep{2009AnGeo..27.3479L} and Self-Similar Expansion (SSE) \citep{2012ApJ...750...23D}. The FP, HM and SSE conversion methods require knowledge, or assumption, of the 3-D propagation direction of a solar wind transient (such as from its source location on the Sun). As we do not know the propagation direction of the CME {\it a priori}, it is easiest to assume that it propagates at an angle of 90~$\degree$ from the Sun-spacecraft line ({\it i.e} in the POS). The POS approximation (as described by Equation~1) is, in fact, a special case of the FP conversion methodology. The PP approximation works on the principle that the maximum brightness contribution along any line-of-sight (LOS) comes from the point closest to the Sun ({\it i.e,} the Thomson sphere), as discussed by, for example, \citet{2006ApJ...642.1216V} and \citet{2012ApJ...752..130H}. \citet{2012ApJ...752..130H} and \citet{2015arXiv151200651I} have shown that, in fact, Thomson scattering maximises over a wide range of angles around the Thomson sphere (the so-called Thomson plateau). We discount the use of PP mainly because it traces out a non radially-propagating point on a CME front. Here, as discussed in section~\ref{polarsect}, we adopt the POS approximation for deriving radial distance from elongation, which is consistent with the original CACTus ethos.\\  
It is worth making the point here that it is actually possible to estimate the 3D propagation angle (and radial speed) of a solar wind feature, such as a CME, without having {\it a priori} knowledge of its source region provided that it can be tracked over sufficient elongation extent. This can be done by analysing the curvature of its associated ridge in the J-map. Such curvature is present even if the feature is propagating at a constant speed in a fixed direction (see, for example, figure 2 of \citet{2012ApJ...750...23D}, where the authors plot simulated time-elongation profiles curves for FP, HM, and SSE geometries). By assuming a fixed geometry, the time-elongation profile of the ridge can be analysed to provide an estimate of the 3D propagation direction and radial speed.
The current version of CACTus can only detect straight lines. Hence we cannot estimate the 3D propagation angle in this way. \citet{2014ApJ...787..119M} have reported that CMEs can decelerate out to 1 AU. Thus, in addition to the geometric artefact discussed above, any deceleration (or indeed acceleration) of a CME during its propagation will modify the curvature due to aforementioned effect. However, it should be noted that, even if some curvature is present, CACTus can still detect a ridge if it is sufficiently bright (such as ridge detected at 20 hours at 15 R$_{\sun}$ in figure~\ref{fig3}). However, any velocity estimate may not be accurate.

\begin{figure*}
\includegraphics[scale=0.55]{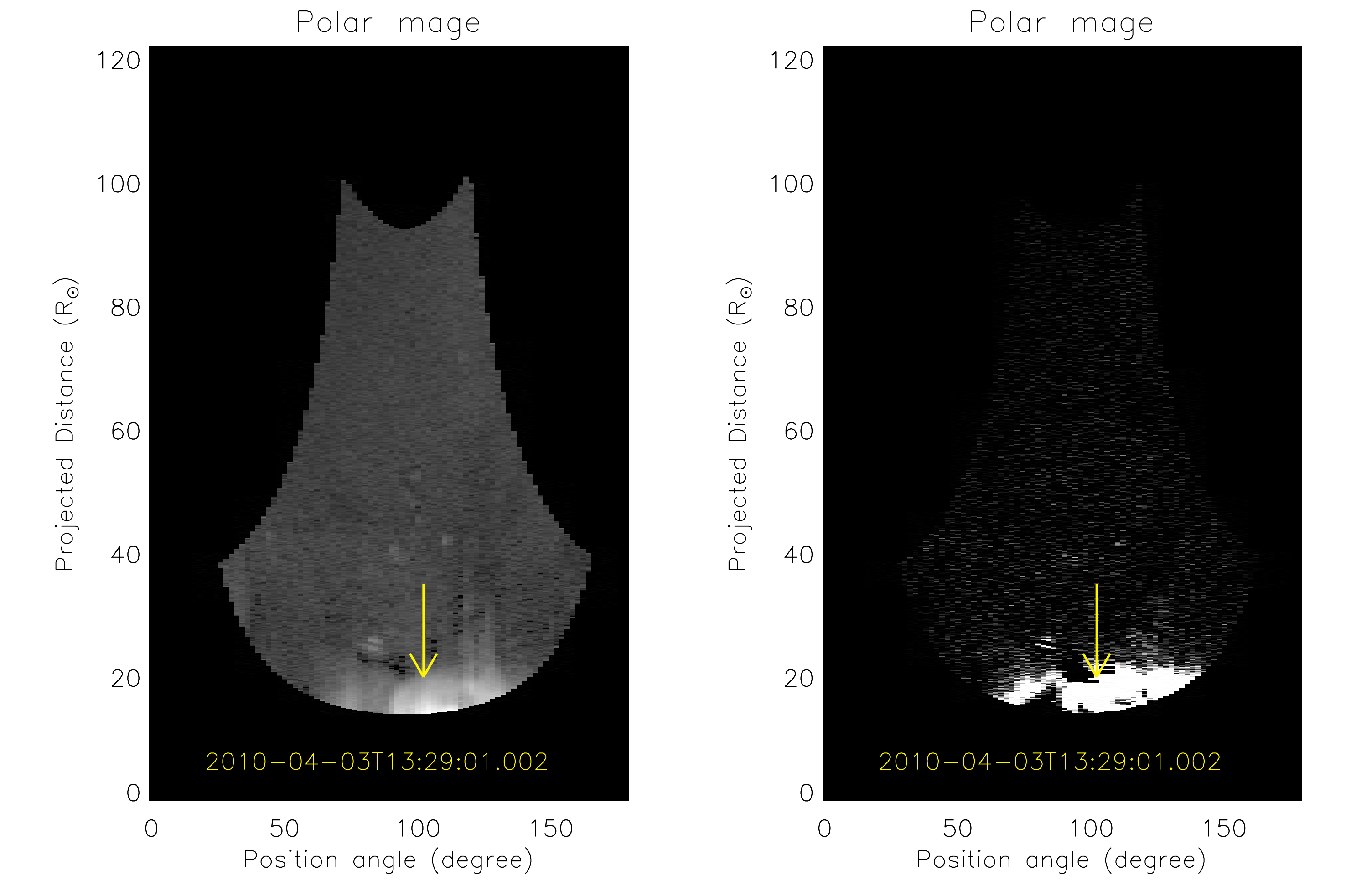}
\caption{\textrm{Left}: Polar transformation of the image presented in the right hand panel of figure.~\ref{fig1}.  \textrm{Right}: Difference and thresholded version of the left hand image. The yellow arrows represent the leading edge of the CME co-spatial with the arrow shown in figure.~\ref{fig1} .}
\label{fig2} 
\end{figure*}

We do not perform CME detection in HI-2 imagery, not least because of the inherent curvature in the signature of a CME propagating over a large range of elongations \citep{2012ApJ...750...23D}. As noted above, such curvature could adversely affect CME detection using the Hough transform implemented in CACTus, as the Hough transform only detects straight lines. However in the case of observations over a limited elongation extent (less than some 15$\degree$), the track of a CME propagating radially outwards at a constant velocity would be virtually straight. In practice, there would still be a some slight curvature over the elongation range covered by HI-1 but not enough to affect the performance of the algorithm, particularly since CMEs are mainly detected from the inner edge of the HI-1 FOV, an elongation of 4 $\degree$ out to 15 $\degree$; the latter equated to a POS distance of 57 R$_{\sun}$. The slightly curved ridge in the left hand panel of figure~\ref{fig3} (detected in HI-1 FOV at 20 hours at 15 R$_{\sun}$ ) is successfully detected by Hough transform (shown in right hand panel of figure~\ref{fig3}) because it is bright. However, the speed estimate in this and similar cases may not be accurate.\\
Difference $\theta-r$ images (similar to the one that is shown in right hand panel of figure~\ref{fig2}) are stacked over time to produce a [$\theta$,$r$,$t$] datacube, where $\theta$ is the PA, $r$ is the radial distance and $t$ is the time in hours from the start of the interval under analysis (up to 72 hours in this case). 

\subsection{Application of the Hough Transform}
\begin{figure*}
\includegraphics[scale=0.55]{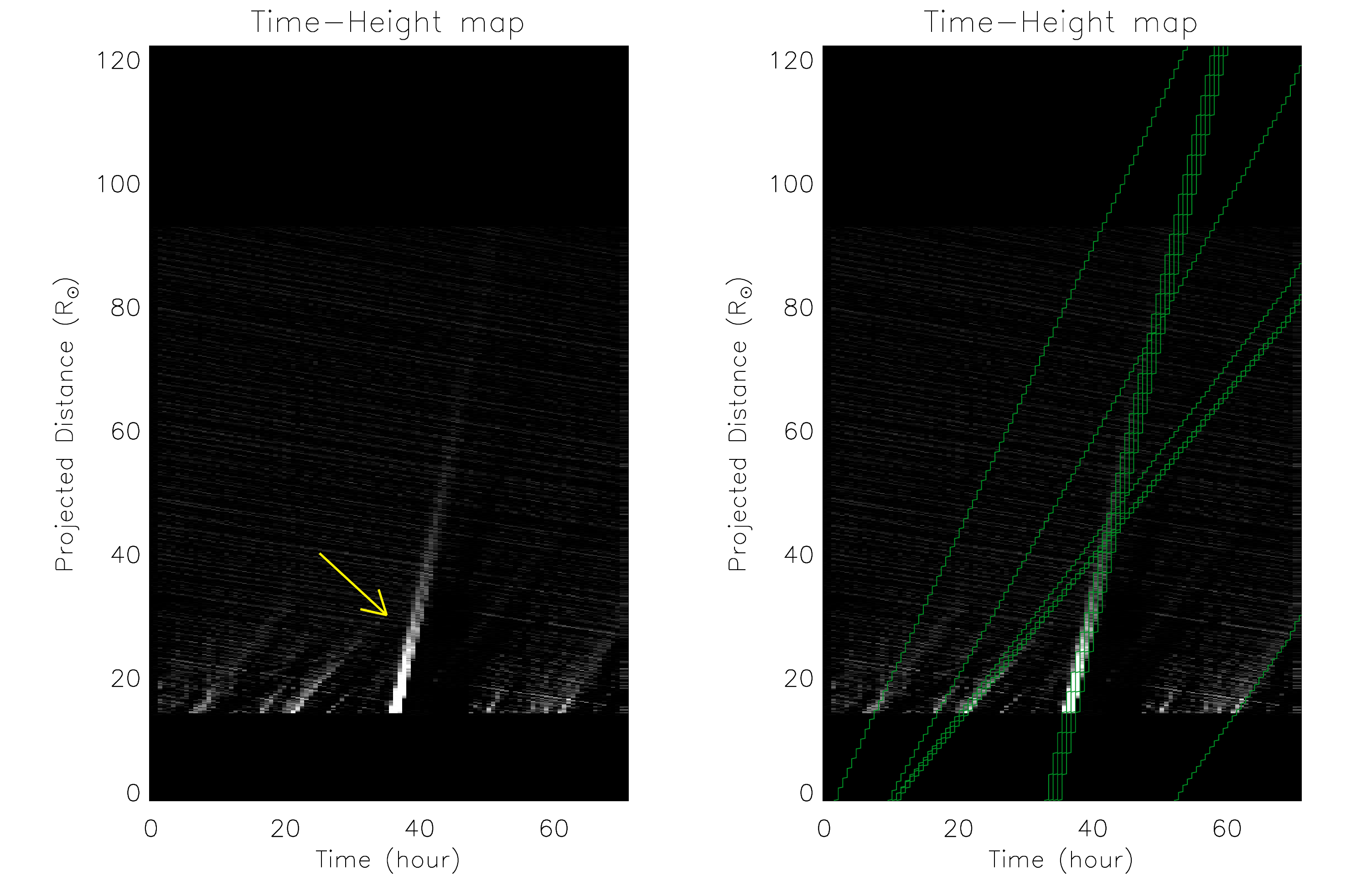}
\caption{\textrm{Left}: Time-height map at 102$\degree$ position angle, generated from HI-1 images from STEREO-A, covering the interval that extends from 2010/04/02 to 2010/04/04. The yellow arrow indicates the inclined ridge corresponding to the leading edge of the particular CME of interest. \textrm{Right}: As left, but overplotted with green curves that represent the significant ridges detected through the application of the Hough transform.}
\label{fig3} 
\end{figure*}

\begin{itemize}
\item For each value of the PA, $\theta$, in the resultant [$\theta$,$r$,$t$] datacube, we have what is effectively a time-height ([$t$,$r$]) map. The left hand panel of figure~\ref{fig3} shows the time-height map at $\theta$~=~102~$\degree$, for the interval covering the propagation of the CME shown in figures~\ref{fig1}  and \ref{fig2} through the FOV of HI-1 on STEREO-A. This PA corresponds to the  the yellow arrows in figures~\ref{fig1}  and \ref{fig2} . The x axis represents the time in hours from 00:00 UT on 2010/04/02 and the y axis represents POS distance from Sun-centre. 
\item Inclined ridges with a positive slope represent features traveling away from the Sun, such as the leading edge of the CME shown in figures~\ref{fig1} and \ref{fig2} (the track of this CME front is indicated by a yellow arrow in figure~\ref{fig3}). Faint ridges with negative slope correspond to the residual star signal. Prior to superior conjunction (since which the STEREO-A spacecraft has been rotated by 180$\degree$), the motion of STEREO-A was such that the apparent drift of the star field across HI images was anti-sunward ({\it i.e} in the opposite direction to CMEs). One should be aware that there could be multiple CMEs propagating along a particular PA over any extended time range (which is why we mark the ridge corresponding to the CME of interest here with a yellow arrow). 
\item The Hough transform is defined as the mapping from image space (in this case a time-height map) to parameter space (or accumulator space) \citep[see, ][]{1997ipsa.book.....J,2004A&A...425.1097R}. An inclined ridge is uniquely characterized by two parameters, its slope ({\it m}) and intercept ({\it c}), provided the slope is not infinite. Since CMEs propagate anti-sunward with finite velocities, the inclined ridges related thereto have positive finite slope. One ridge maps to a single point in accumulator space \citep[see, figure 2 in][]{2004A&A...425.1097R}. The weight given to the point in  accumulator space is determined by the number of points lying along the ridge in the time-height map. Brighter ridges in the time-height map tend to be given larger weights in accumulator space than fainter ridges, as they can be usually tracked out further in radial distance. 
\item As noted above, CACTus uses the Hough transform to isolate (as straight lines) the significant ridges in time-height maps associated with the passage of CMEs.  We filter out those points in accumulator space with a low weighting by applying a weight threshold, $W_{thresh}$ that is given by,
\begin{equation}
W_{thresh}=W_{mean} + f \times W_{sd}
\end{equation}
where, $W_{mean}$ and  $W_{sd}$ are the mean and standard deviation of the weights of all points in accumulator space and $f$ is an arbitrary factor. CACTus uses an empirically-based value of $f$ = 4 in the analysis of the STEREO/HI-1 data, based on analysing a sample of CMEs.
\item Points in  accumulator space with weights that exceed $W_{thresh}$ are taken to be the points that correspond to the tracks of CMEs in time-height maps. The time-height map presented in the right hand panel of figure~\ref{fig3} reproduces that shown in the left hand panel, but with significant ridges (corresponding to those points in  accumulator  space that exceed $W_{thresh}$) overplotted in green. We plot the green curves over the full height range. The coordinates of the detected points in  accumulator  space, {\it i.e} {\it m} and {\it c} of each detected ridge in the time-height map, provide an estimate of the POS velocity ($v$) and time of first appearance of each detected CME ($t_{0}$), respectively. $t_{0}$ does not represent the time that the CME first appears in the HI-1 FOV but, instead, corresponds to the back-projected time at which the CME would be at Sun-centre ({\it i.e,} $r$=0). This is because the datacube input into CACTus extends from $r$=0. To estimate the time at which the CME enters the HI-1 FOV, a correction to t$_{0}$ is required; no such correction to the velocity is required. The correction depends on the velocity of a ridge and the distance to the inner edge of the HI-1 FOV. Unlike coronagraph images where the inner edge of the FOV is independent of $\theta$ (because the occulter is circular), the POS distance of the inner edge of the HI-1 FOV depends on $\theta$ (see figure~\ref{fig2}). For the time-height map presented in figure~\ref{fig3}, this distance is 15 R$_{\sun}$. The $\theta$ dependent correction of $t_{0}$ that is implemented in the current version of CACTus was not present in earlier versions. As discussed above, a ridge in time-height map corresponds to the propagation of a CME along a particular PA ($\theta$). We perform the Hough transform on time-height maps at every PA within the HI-1 FOV. Since CMEs are extended in $\theta$, a single CME will correspond to a cluster of points in the resultant [$\theta$,$v$,$t_{0}$] datacube. 
\item CACTus subsequently performs integration over $v$ in order to obtain a [$\theta$,$t_{0}$] map. We use a morphological closing technique to fill the gaps between the points of a cluster in  such a map. This technique performs dilation of an image (as described above) followed by the contraction using a kernel of a given size \citep[see, ][]{2002dip..book.....G}. This process fills gaps between the points in a cluster in the [$\theta$,$t_{0}$] map that are smaller than the size of kernel. 
\item Unlike the version of CACTus implemented on coronagraph images, which uses a one dimensional kernel with a width of 5 bins in the $\theta$ direction, a two dimensional kernel is used for detecting CMEs in HI-1 data. This kernel has a width of 8 bins in $\theta$ and 5 bins in $t_{0}$. This means that CACTus bridges the gaps between the points in accumulator space that are less than 16 $\degree$ in $\theta$ and 200 minutes in $t_{0}$. These values are empirically chosen (through examination of a number of CMEs observed by HI-1) such that they connect different parts of the same CME together without (too often) combining different CMEs. Such morphological closing is required because, for example, a single CME often comprises more than one distinct feature (such as a leading edge and prominence material) and, moreover, the brightness of even a single feature may be non-uniform in $\theta$. 
\item The left panel of figure~\ref{fig4} shows a [$\theta$,$t_{0}$] map, which we will henceforth refer to as a CME map, that spans the same interval as presented in figure~\ref{fig3} ({\it i.e,} 2010/04/02 to 2010/04/04). The x axis represents position angle and y axis, time. The CME map illustrated in the right hand panel of figure~\ref{fig4} presents the location, as a function of both PA and time, of the maximum POS velocity for each of the 5 CMEs detected by CACTus during the time interval covered by the time-height maps presented in figure~\ref{fig3}. The magnitude of the velocity is colour coded according to the colour bar. The white contours overplotted  on the right hand panel of figure~\ref{fig4} illustrate the identified boundary of each cluster. Each cluster, of course, represents a separate CME detected by CACTus. Cluster 4 in the left hand panel of figure~\ref{fig4} corresponds to the CME under particular consideration here, images of which are shown in figures~\ref{fig1} and \ref{fig2}.
\end{itemize}
It is also important to note that, for various reasons, there are occasional data gaps in the HI-1 telemetry stream. This not only leads to missing images, but also occasional incomplete images (so-called ``missing blocks''). One advantage of the Hough transform is that it can detect ridges even in the presence of data gaps, provided the ridge a sufficiently bright and the data gap is sufficiently short (so that the weight of the point in Hough  space corresponding to the ridge still exceeds W$_{thresh}$). Since we apply morphological closing in $\theta$ and time, data gaps will not adversely affect the results unless they are too long or too frequent. If there are too many data gaps, CACTus might not detect any ridges or may detect multiple ridges. While this could be resolved by resampling prior to implementing the Hough transform, there are too few data gaps in the HI-1 imagery for this to be necessary.

\begin{figure*}
\includegraphics[scale=0.55]{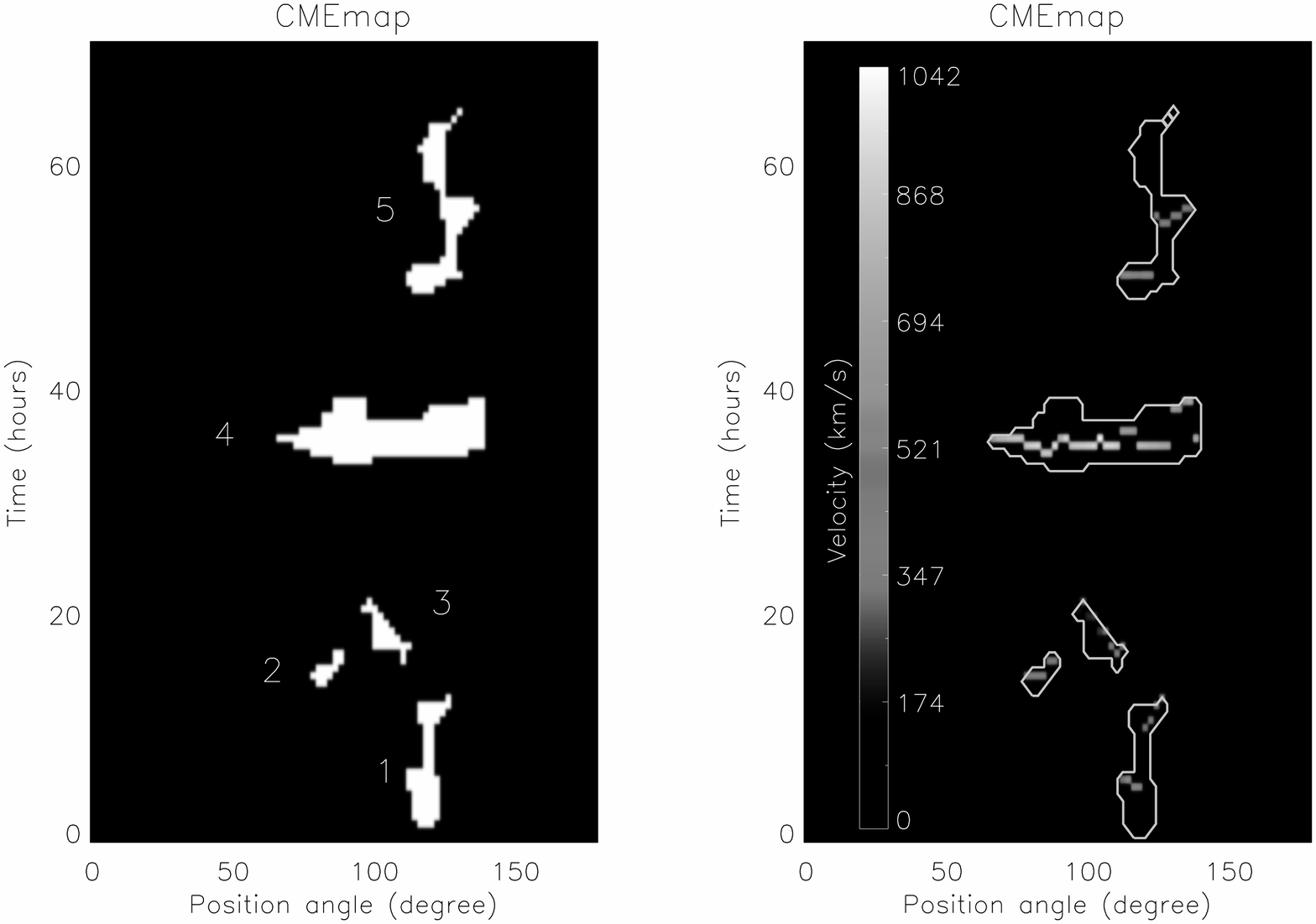}
\caption{\textrm{Left}: CME map showing 5 clusters of points representing 5 different CMEs detected during the time interval extending from 2010/04/02 to 2010/04/04. Clusters are numbered according to their time of first appearance. \textrm{Right}: CME map showing for each cluster (i.e, for each CME), the maximum velocity (color coded) at each PA over the entire PA extent of the CME. White contours represent the identified boundaries of the clusters.}
\label{fig4} 
\end{figure*}
\subsection{Determination of CME position angle (PA) width, time of appearance and velocity}
\label{estimateparam}
\begin{figure*}
\centering
\includegraphics[scale=0.43]{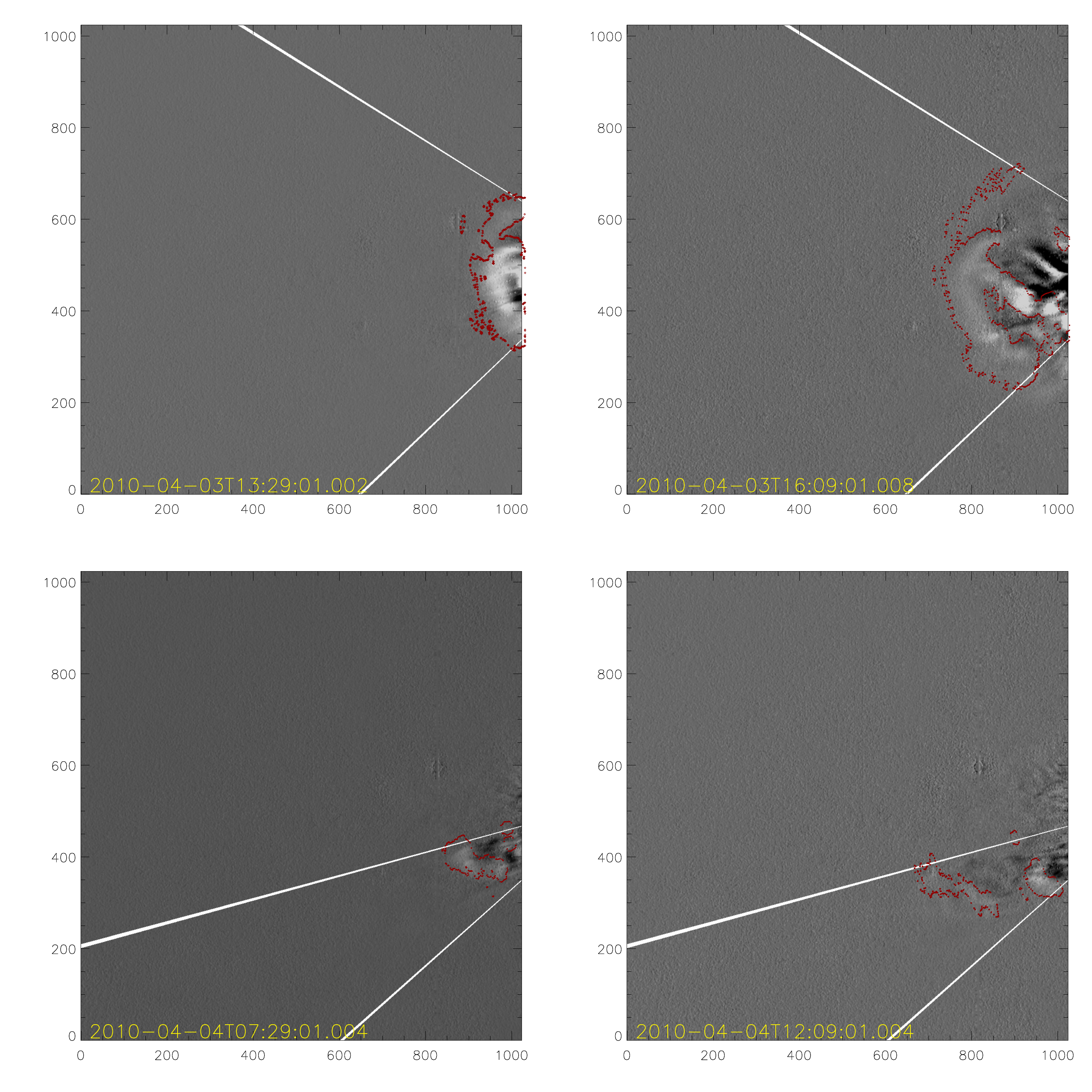}
\caption{\textrm{Top}: Difference images of the CME corresponding to cluster 4 in the left hand panel of  figure.~\ref{fig4} (No 4 in table~\ref{table1}). White lines delimit the northernmost and southernmost PA extents of the CME. \textrm{Bottom}: Same as left hand panel but for cluster 5 (No 5 in table~\ref{table1}). Points overplotted in red represent the perimeter of the radially outward moving features identified by application of Hough transform. }
\label{fig5} 
\end{figure*}

\begin{figure*}
\subfigure{\includegraphics[angle=0,scale=.47]{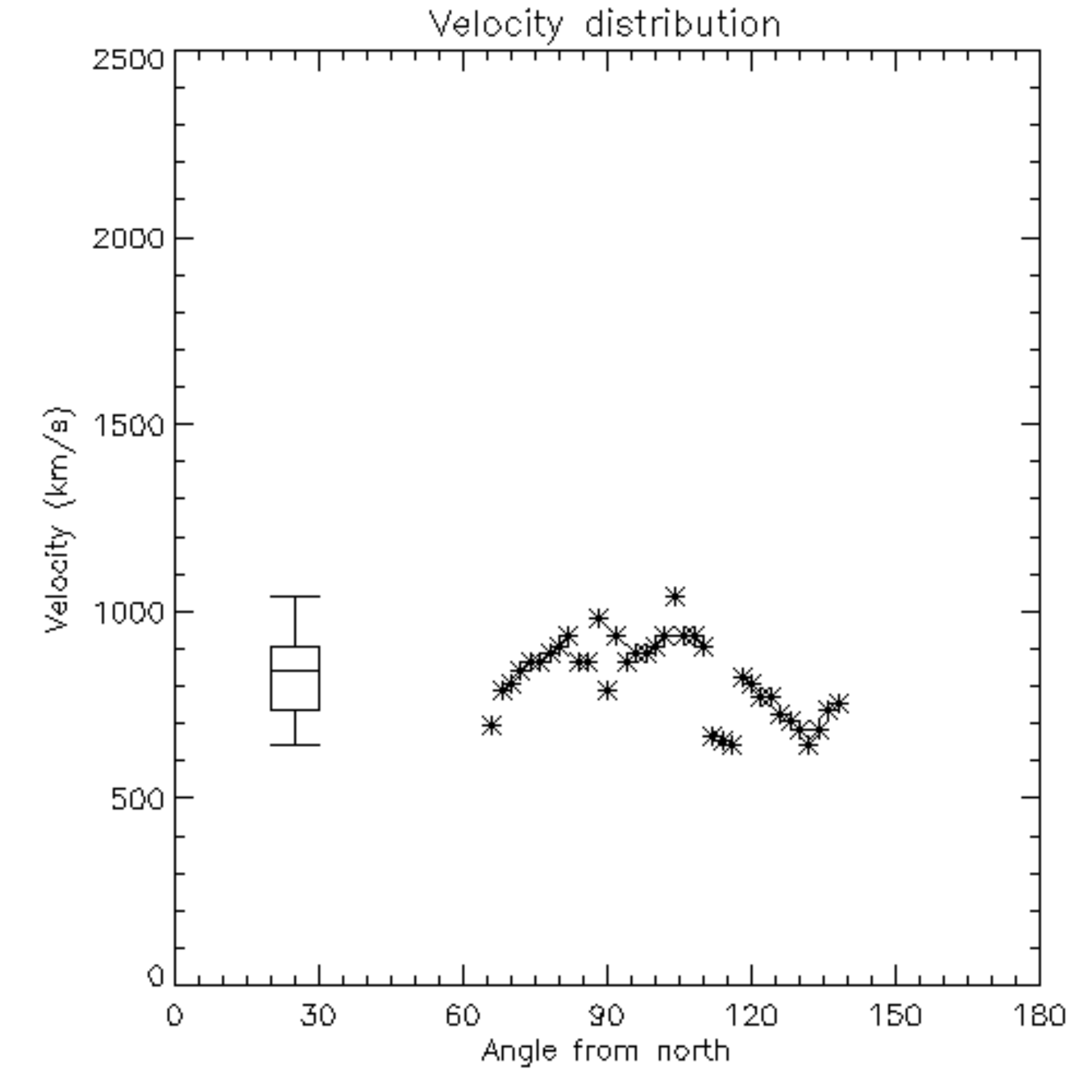}}
\subfigure{\includegraphics[angle=0,scale=.47]{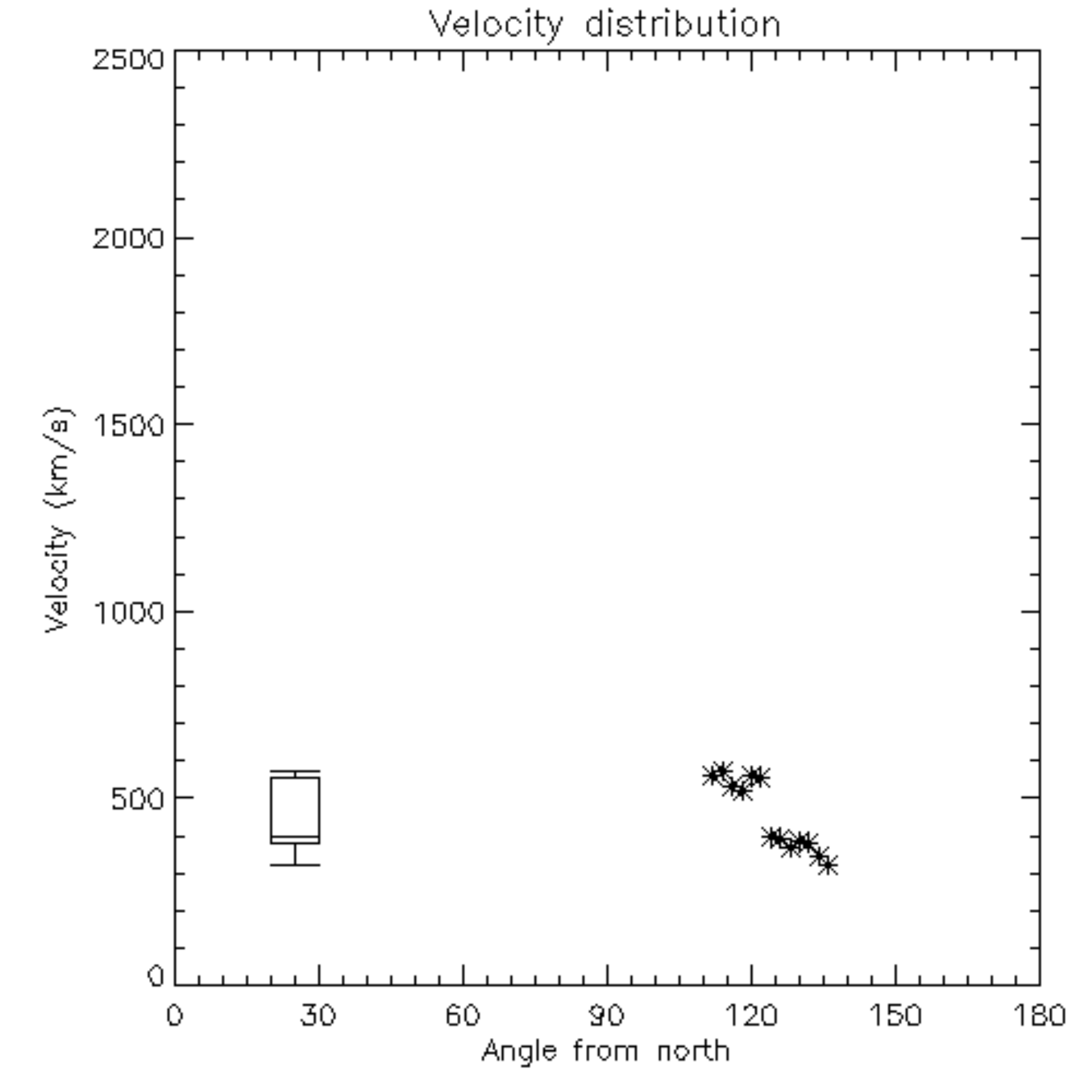}}
\caption{\textrm{Left}: Maximum velocity as a function of position angle (measured counterclockwise from solar north) for the CME detected as cluster 4 in the left hand panel of figure~\ref{fig4}. \textrm{Right}: Same as left, but for cluster 5. The box and whisker keys indicate the median and quartiles of the velocity distribution, as well as the minimum and maximum velocity values (see text for details).}
\label{fig6} 
\end{figure*}

\begin{itemize}
\item \textbf{Estimation of PA width:} Once the so-called CME map has been created, the PA width of a CME is estimated by calculating the extent of the corresponding cluster in the CME map in the $\theta$ direction. It is important to note that, for each CME the PA width provided by CACTus is the maximum width of that CME throughout its propagation through the HI-1 FOV. The central position angle (CPA) of propagation of a CME is calculated as the mid-point of its associated cluster in $\theta$. Figure~\ref{fig5} shows difference images for two of the CMEs (CMEs that correspond to clusters 4 and 5 in figure~\ref{fig4}) that are detected by the application of CACTus to HI-1 imagery from STEREO-A during the time interval extending from 2010/04/02 to 2010/04/04 (see animated figures~\ref{fig7} and \ref{fig8} available online). In each case the PA width yielded by the method is delimited by white lines. 
\item \textbf{Estimation of time of appearance:} To estimate the time of appearance of a specific CME in the HI-1 FOV, a local background intensity and standard deviation for the corresponding cluster in the CMEmap is first estimated using the [$\theta,r,t$] datacube. For each cluster in a CMEmap, a local background intensity is estimated by fitting a straight line to the intensity that is summed over the PA width and radial distance extent of the [$\theta,r,t$] datacube over a time range that extends from 30 images before the lower boundary of a cluster to 30 images after the upper boundary of the same cluster. The value of 30 is chosen empirically. The time of appearance of a CME is defined as being the time beyond/after the lower boundary of a cluster at which the intensity first exceeds the background value by 2$\sigma$, where $\sigma$ is the standard deviation of the intensity. We do not simply use the lower boundary to determine the time of appearance as CMEs can generate waves or shocks that can be manifest as regions of faintly enhanced brightness prior to the appearance of the CME itself. Moreover, small blobs are sometimes observed to propagate radially outward ahead of a CME \citep{2009ApJ...691.1222R}. Such brightenings ahead of CMEs may (or indeed may not) be detected by the Hough transform. If such features are identified in accumulator space, they may subsequently be amalgamated with the main cluster after the application of the morphological closing technique. Therefore, the lower boundary itself may not represent the true time of CME appearance. If any such feature or unstructured flow (for example, streamer blobs) prior to CME appearance is bright enough to exceed the 2$\sigma$ threshold then it will affect the estimated time of arrival (potentially making it earlier than what would be deduced from visual inspection).
\item \textbf{Estimation of velocity:} Each bin in a CME map that falls within a CME cluster is associated with a velocity, v. For each cluster of points in a CME map, and thus for each CME, we identify the maximum velocity at each PA within the identified CME PA span (right hand panel of figure~\ref{fig4}). The left hand and right hand panels of figure~\ref{fig6} show the maximum velocity as a function of PA for the CMEs that correspond to clusters 4 and 5, respectively, in figure~\ref{fig4} (images of which are presented in figure~\ref{fig5}). Overplotted on each panel is a box-and-whisker key depicting the range of maximum velocities over the entire PA extent of the corresponding CME. The upper and lower boundaries of the box itself represent the upper and lower quartiles of the distribution of maximum velocities; the median value is marked by horizontal line inside the box. The whiskers encompass the highest and the lowest value of the maximum velocity.
\end{itemize}
Estimated parameters of all 5 CMEs that were detected by CACTus in HI-1 images from STEREO-A/HI-1 images during the time period extending from 2010/04/02 to 2010/04/04 (including the two CMEs shown in figure~\ref{fig5}) are listed in table~\ref{table1}. 


\section{Comparison with manual catalog}
\label{mancatalog}

We have applied CACTus to eight different days of STEREO-A/HI-1 images (2010/04/04, 2010/04/03, 2010/04/02, 2008/12/12, 2008/04/26, 2008/02/04, 2007/10/22 and 2007/04/19) and have compared the results with those of manual CME cataloging endeavors, in particular those performed as part of the EU FP7 Heliospheric Cataloging Analysis and Techniques Service (HELCATS) project \footnote{\url{http://www.helcats-fp7.eu}}. The manual catalog compiled as part of HELCATS work package 2 contains observational CME parameters (in particular, time of first observation and northern and southernmost position angle extents) obtained through visual inspection of the STEREO/HI-1 images. This catalog is available online \footnote{\url{http://www.helcats-fp7.eu/catalogues/wp2_cat.html}}. Indeed, the application of CACTus to the STEREO/HI-1 imagery documented here is also performed under the auspices of the HELCATS project. The HELCATS manual catalog is described, and preliminary analysis undertaken, by \citet{harrison2016}. Moreover, we compare the CACTus POS speed estimates with those contained within the augmented (work package 3) version of the HELCATS manual CME catalog \footnote{\url{http://www.helcats-fp7.eu/catalogues/wp3_cat.html}} that also includes estimates of  3D radial speeds calculated along, or at least near, the CPA of each CME. The 3D speeds in this augmented catalog are derived through the analysis of the time-elongation profiles of the CMEs manually tracked in time-elongation maps (J-maps) generated from combined HI-1 and HI-2 imagery, using three techniques : Fixed-$\phi$ Fitting \citep[FPF;][]{2008GeoRL..3510110R}, Self-Similar Expansion Fitting (SSEF) with an assumed angular half width of 30 degrees \citep{2012ApJ...750...23D} and Harmonic Mean Fitting \citep[HMF; ][]{2009AnGeo..27.3479L}. The FP, HM and SSE geometries that underlie these fitting techniques assume different values for the cross sectional width of a CME. FP assumes a half width of 0 $\degree$, {\it i.e,} CME is a point source whereas, HM assumes a half width of 90 $\degree$, {\it i.e,} a circle tied to Sun-center. The SSE geometry is more general, with a half width that can be anywhere between 0 to 90 $\degree$. 30 $\degree$ is chosen for use in the HELCATS manual catalogs to account for the fact that the average angular span of  CME is around 60$\degree$.
We reiterate here that the POS approximation used by CACTus is a special case of the FP approximation, assuming CME propagation to be at 90 $\degree$ from the Sun-spacecraft line.
\subsection{Comparison of number of events detected}
During the 3 day period extending from the beginning of 2010/04/02 to the end of 2010/04/04, CACTus detected 5 CMEs (see, figure.~\ref{fig4} and tables~\ref{table1} and \ref{table3}). Since we incorporated that three day period into a single CACTus analysis, we get only a single CME map (shown in figure.~\ref{fig4}) that contains all events detected during that period.  The other 5 days were analysed individually. Table.~\ref{table3} compares the number of CMEs detected by CACTus with the number detected manually as a part of the HELCATS project. We find that CACTus detected more events than were listed in the manual catalog during the 8 days analysed (11 as opposed to 6). Over-estimation of the number of CMEs appears to be a common feature of CACTus \citep[see][]{2009ApJ...691.1222R}. Having examined the CMEs automatically identified in the HI-1 data, we find that, in some cases, CACTus identifies multiple separate events that, on closer investigation, are part of a single CME. {\it For example,} one event detected by CACTus on 2008/04/26 was found to actually be the flank of a separately-detected  CME. In that case, both the apex and the flank of the CME were bright while the region between them was faint, hence their detection by CACTus as two separate events. The CME identified by CACTus on 2010/04/04 (figure~\ref{fig5} (bottom panel)) is not listed in the HELCATS manual CME catalog. This is probably because this CME is narrower than the 20$\degree$ PA width threshold imposed in the HELCATS manual cataloging. Similarly, the three events detected by CACTus on 2010/04/02 (CMEs 1, 2 and 3 in figure~\ref{fig4}), all are narrow unstructured features propagating radially outward (these are still identified as CMEs by CACTus according to the definition implemented therein) and are therefore, not listed in manual HELCATS catalog. In this work, we have set a threshold of 10$\degree$ in PA width for a feature detected by CACTus to qualify as CME; application of CACTus to coronagraph images generally uses 5$\degree$. 

In summary CACTus detected all CMEs present in the manual catalog although one of them was erroneously identified by CACTus as two separate events. Moreover, use of a 10$\degree$ (rather than 20$\degree$) PA width threshold resulted, over these 8 days, in the identification by CACTus of four additional, narrow, CMEs. It could be argued that such narrow CMEs which (presumably they are small blobs of plasma) usually follow a larger CME ought not to be categorised as a separate CME but as a part of the preceding CME. In manual cataloging, such association is at the discretion of the operators. CACTus applies a more objective definition; any feature (with brightness exceeding W$_{thresh}$ ) that appears more than 200 minutes after the appearance of a CME is considered as a separate CME. Although this is the objective application of what is still a subjective threshold, it means that CME detection will remain consistent over a long period of time provided that there are no significant changes in instrument performance \citep[see,][]{2016SoPh}. \\
\subsection{Comparison of angular width}
We also note that CACTus estimates consistently lower PA widths for the CMEs detected during the days under study than are presented in the HELCATS manual catalog (see, table.~\ref{table2}). We suggest that a possible reason for this is that CMEs expand as they propagate outward. Moreover, the flanks of a CME are, often fainter than its leading edge. Consequently, the ridge corresponding to the flank of a CME in a time-height map at a given PA may not be sufficiently extended or bright to result in its detection by CACTus ({\it i.e.} its weight may not exceed $W_{thresh}$). 
Another possible reason for the discrepancy in manually and automatically-deduced PA widths could be related to the fact that for fast CMEs in particular, the CME is often surrounded by a shock that tends to be fainter than the CME itself. We believe that it is more likely that a manual cataloger would be influenced by the presence of even faint shocks when estimating the CME width. Depending upon the brightness of the shock, it may (or may not) be detected by CACTus. It is debatable whether a shock or any structures adjacent to the CME that are deflected due to the presence of a shock should (or should not) be included when estimating the width of a CME. However, we believe that the CACTus results would only be affected by the presence of the brightest shocks ({\it i.e} those with the longest tracks in a time-height map). It should be pointed out that the discrepancy in the PA width estimation is also seen when comparing the width of CMEs detected by CACTus in coronagraph data with those detected manually \citep{2009ApJ...691.1222R}. Thus we believe that the estimation of the PA width of CMEs in HI-1 data is affected by: (a) possible non-radial motion of CMEs, (b) the tendency of CME flanks to be faint and hence their ridges in time-height maps to be of limited extent and (c) the presence of faint shocks around CME flanks.
Potentially the PA widths estimated by CACTus could be made more consistent with the values in the manual catalog by lowering the intensity threshold, but doing so would also increase the number of false detections.
\subsection{Comparison of time of appearance in the HI-1 FOV}
\label{timecomparison}
We find that the times of first appearance ($t_{0}$) of the CMEs in the HI FOV, as estimated by CACTus and in the HELCATS manual catalog, are in fairly good agreement (see, table.~\ref{table2}), with the exception of the event on 2007/04/19. For that event, CACTus yields a time of entry into the HI-1 FOV that is two hours earlier than the value quoted in the manual catalog. Having examined the images, we believe CACTus to be correct; the CME is already well inside the HI-1 FOV at the time of first appearance quoted in the HELCATS catalog. The reason for this discrepancy is not clear. Nevertheless, it confirms the rather subjective nature of manual cataloging. In general, it is thought that such time differences are due to a) narrow blobs preceding a CME \citep{2009ApJ...691.1222R} or b) the inclusion of the shock ahead of a CME. This is discussed in greater detail in section~\ref{sec4}.

\subsection{Velocity comparison}
We have compared the POS velocity from CACTus with the best-fit 3-D velocity derived using FPF (v$_{FP}$), SSEF (v$_{SSE}$) with an assumed angular half width of 30 $\degree$, and HMF (v$_{HM}$). The speeds are quoted in table~\ref{table2}. $\phi_{SSE}$ also listed in table~\ref{table2} refers to the best-fit angle between the Sun-observer line (in this case STEREO-A) and the propagation direction of the CME returned by the SSEF technique ($\phi_{FP}$ and $\phi_{HM}$ are also quoted in HELCATS catalog, and for the CMEs considered here are similar in value to $\phi_{SSE}$). Since SSE assumes a more generalized geometry which reduces to FP and HM geometries under certain limits, we have quoted only $\phi_{SSE}$  for the CMEs considered in table~\ref{table2}. In particular we compare the POS velocities from CACTus with those derived using the FPF technique, because the POS geometry is a special case of the FP geometry.  

Note that $\phi_{SSE}$,  is a reasonable proxy for the longitudinal separation between the CMEs and the spacecraft.
We find that the POS speed from CACTus and v$_{SSE}$ match reasonably well except for the CME on 2007/10/22. We note that $\phi_{SSE}$ for this event is 174$\degree$, meaning it propagated more or less directly away from the spacecraft. Therefore, its projected speed in the POS would be a gross underestimate of its 3D radial speed. The POS approximation underestimates the 3D propagation velocity of any CME that is propagating in a direction well away from the limb ({\it eg,} a halo CME). For all other events we find that $\phi_{SSE}$ is close to 90$\degree$ (i.e, the CMEs are near-limb events) so it is not surprising that the POS speed is more consistent with the 3D speed derived using the FPF, HMF and SSEF techniques. We also find that, for these near-limb events, the velocities derived using FPF, HMF and SSEF (which are based on three different underpinning geometries) are in fairly good agreement. Remember, for a limb event, the FP geometry is equivalent to the POS approximation. It is worth repeating here that, with the current implementation of CACTus we do not use any {\it a priori} information regarding a CME's propagation angle and we cannot estimate the propagation direction in the same way as FPF, SSEF and HMF techniques as we cannot detect any curvature of the ridges in a time-height map. However the POS approximation is a good approximation for near limb events.

\begin{table}
\begin{center}
\caption{Extract of the CACTus CME catalog for 2010/04/02 - 2010/04/04 }  

\label{table1}
\begin{tabular}{l c l l l l l l l l}
\hline \hline
No&t$_{0}$ & CPA & da & NoPA & SuPA & v & dv & minv & maxv \\
\hline
1&2010/04/02 05:29 & 116 & 12 & 110 & 122 & 459 & 73 & 316 & 498\\
2&2010/04/02 14:09 & 83 & 10 & 78 & 88 & 412 & 25 & 384 & 454\\
3&2010/04/02 18:09 & 104 & 16 & 96 & 112 & 319 & 65 & 229 & 397\\
4&2010/04/03 12:09 & 102 & 72 & 66 & 138 & 823 & 108 & 571 & 1041\\
5&2010/04/04 00:49 & 124 & 24 & 112 & 136 & 397 & 86 & 350 & 571\\
\hline
\end{tabular}
\end{center}
\tablecomments{No represents the cluster number (CME number), t$_{0}$ is the time of appearance of the CME in the HI-1 FOV, CPA is the central position angle (PA) of the CME (degrees), da is its PA width (degrees), NoPA is its northernmost PA extent (degrees), SuPA is its southernmost PA extent (degrees), v is the median of the maximum value of POS velocity at each PA of the CME (km~s$^{-1}$), dv is the (1 $\sigma$) variation of the velocity over the entire PA extent of the CME, minv is lowest value of the maximum velocity and maxv is highest value of the maximum velocity.}
\end{table}



\begin{table}
\centering
\caption{Comparison of the number of events}  

\label{table3}
\begin{tabular}{cc@{		}c}
\hline \hline
Date & Events detected by CACTus &    Events detected in Manual catalog\\
\hline
2010/04/04 & 1 & 0\\
2010/04/03 & 1 & 1\\
2010/04/02 & 3 & 0\\
2008/12/12 & 1 & 1\\
2008/04/26 & 2 & 1\\
2008/02/04 & 1 & 1\\
2007/10/22 & 1 & 1\\
2007/04/19 & 1 & 1\\
\hline
\end{tabular}
\end{table}

\begin{table}
\begin{center}
\caption{Comparison of CACTus vs manual detection}  

\label{table2}
\begin{tabular}{c l l l | c l l l l l l}
\hline \hline
\multicolumn{4}{c}{CACTus} & \multicolumn{7}{c}{Manual} \\
t$_{0}$ & CPA & da & v  &  t$_{0}$ & CPA & da & v$_{SSE}$ & $\phi_{SSE}$ & v$_{FP}$ & v$_{HM}$ \\
\hline
2010/04/03 12:09 & 102  & 72 & 843 & 2010/04/03 12:09 & 102 & 105 & 927 & 77 & 889 &962\\
2008/12/12 15:29& 77& 70& 469& 2008/12/12 15:29 & 77 & 95 & 426 & 69 & 419 & 431\\
2008/04/26 18:09& 82& 72& 595& 2008/04/26 18:49 & 85 & 110 & 649 & 95 & 620 & 679\\
2008/02/04 14:09& 83& 66& 448& 2008/02/04 14:09 & 77 & 115 & 511 & 53 & 506 & 515\\
2007/10/22 14:49& 84& 68& 225& 2007/10/22 14:49 & 77 & 85 & 691 & 174 & 499 & 748\\
2007/04/19 11:30& 90& 72& 376& 2007/04/19 13:30 & 90 & 100 & 392 & 61 & 389 & 393\\

\hline
\end{tabular}
\end{center}
\tablecomments{v$_{SSE}$ is the speed in km s$^{-1}$ and $\phi_{SSE}$ is the spacecraft-Sun-CME angle in degrees derived using SSEF \citep{2012ApJ...750...23D}. If $\phi_{SSE}$ is close to 90$\degree$ then the CME originated from near the limb. If $\phi_{SSE}$ is close to 0$\degree$ or 180$\degree$ then CME is a front or backsided halo, respectively, as seen from the STEREO-A spacecraft. v$_{FP}$ and v$_{HM}$ are the speed in km s$^{-1}$ derived using FPF and HMF, respectively. }
\end{table}


\section{ Online automated catalog of SECCHI/HI-1 generated by CACTus} 
\label{sec4}
\begin{figure*}[h]
\centering
\includegraphics[scale=0.3,angle=-90]{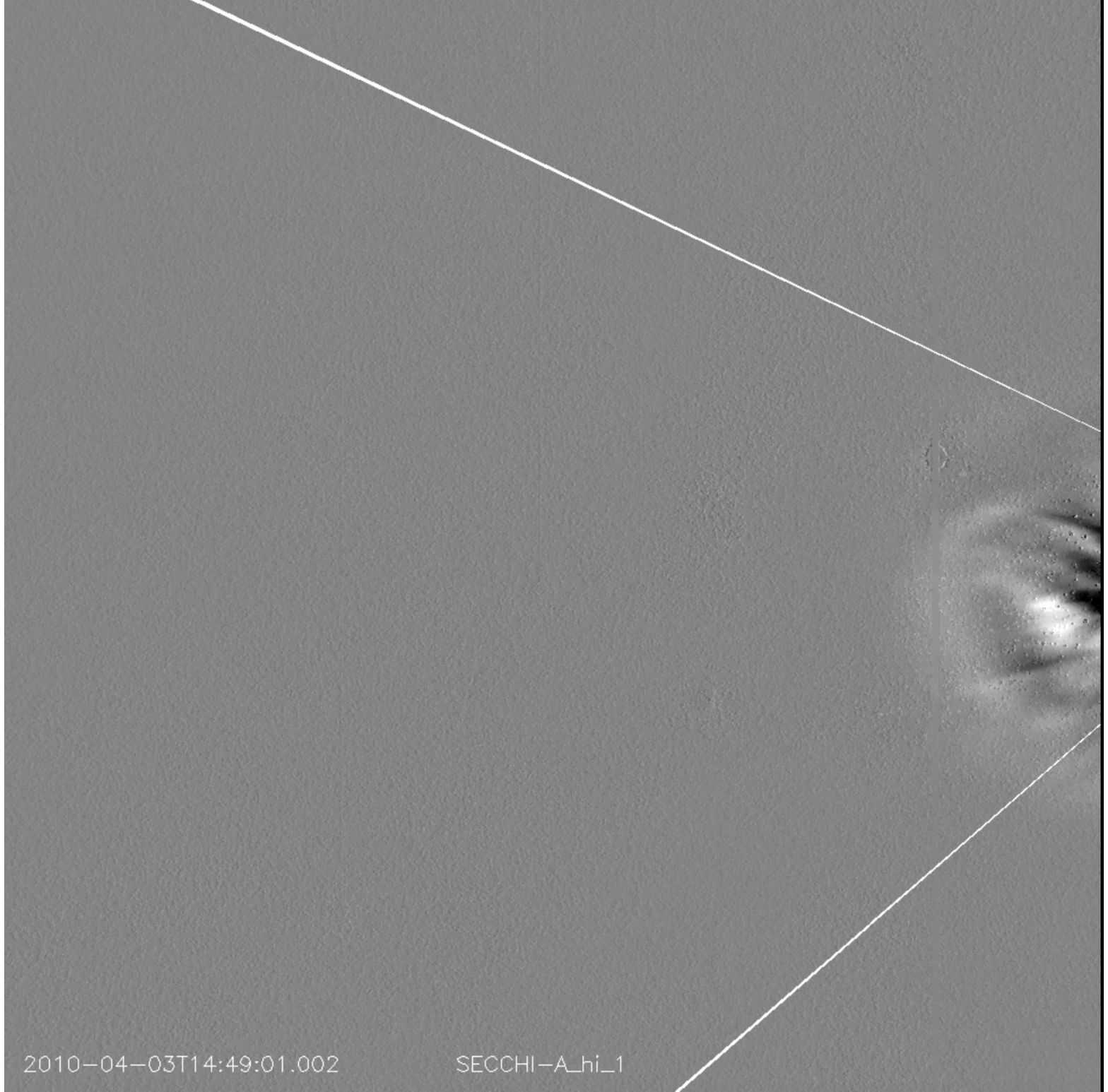}
\caption{Movie corresponding to this animated figure shows the CME moving outward anti-sunward. The angular width of CME is delimited by white lines. }
\label{fig7} 
\end{figure*}

\begin{figure*}[h]
\centering
\includegraphics[scale=0.28,angle=-90]{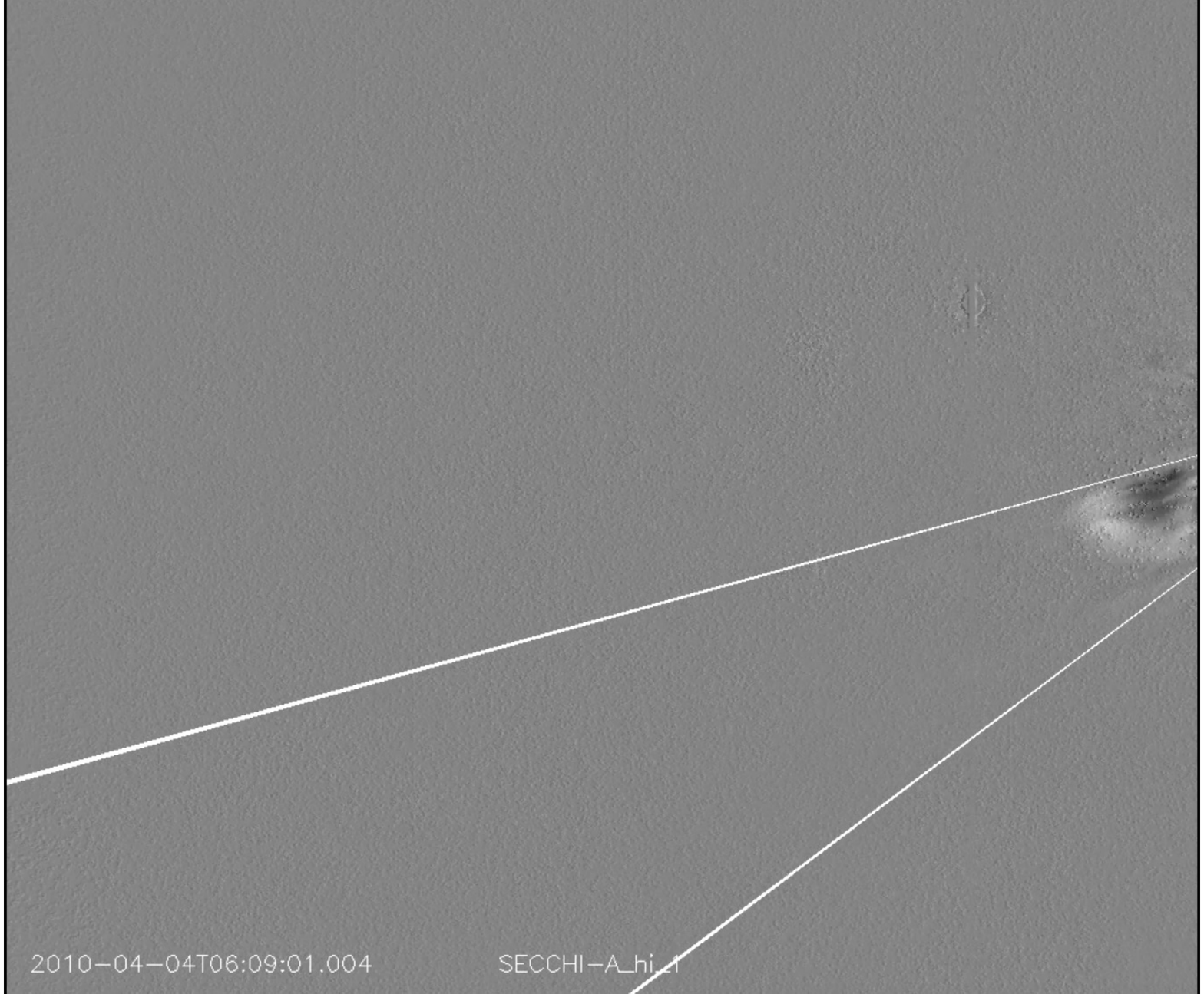}
\caption{Movie corresponding to this animated figure shows a narrow CME moving outward. The angular width of CME is delimited by white lines. }
\label{fig8} 
\end{figure*}

\begin{figure*}[h]
\centering
\includegraphics[scale=0.35]{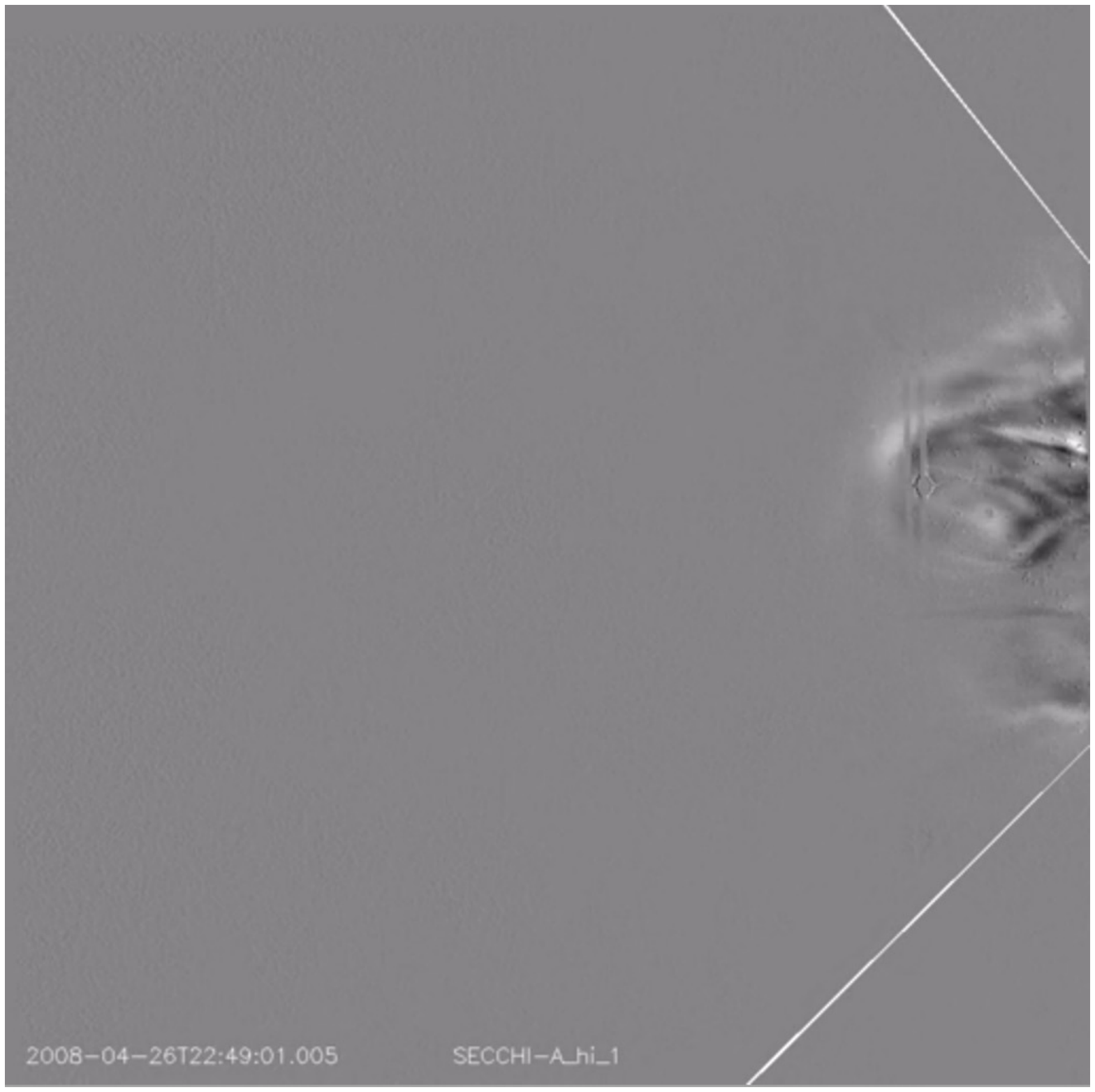}
\vspace*{-2cm}
\caption{Movie corresponding to this animated figure shows the CME moving outward. Pre-CME outflows are seen clearly that affect the estimation of the time of appearance in HI-1 FOV.}
\label{fig9} 
\end{figure*}

\begin{figure*}[h]
\centering
\includegraphics[scale=0.35]{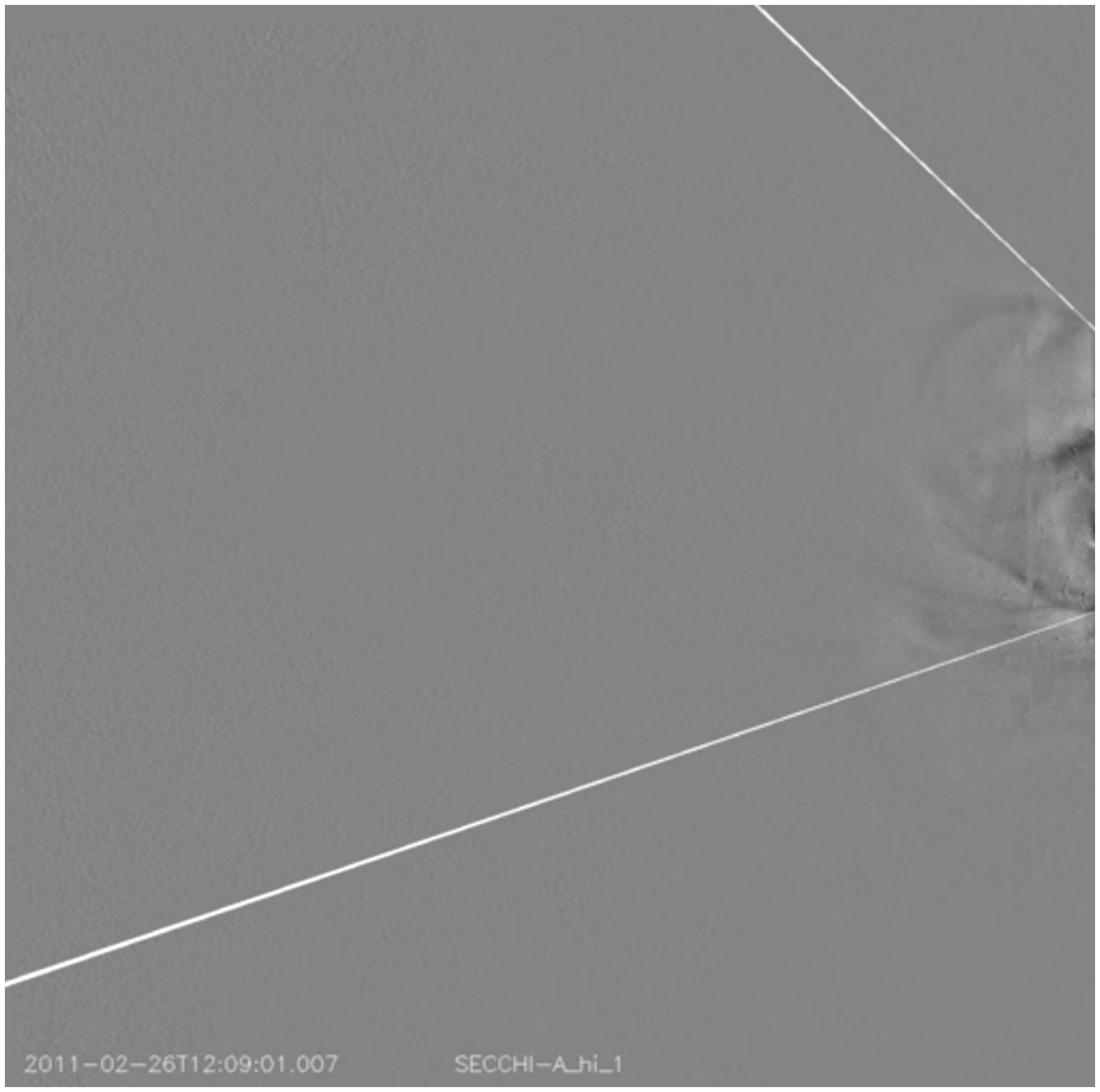}
\vspace*{-2cm}
\caption{Movie corresponding to this animated figure shows a CME with faint leading edge moving outward. The faint and unstructured leading edge makes the estimation of time of appearance of a CME in HI-1 FOV difficult.}
\label{fig10} 
\end{figure*}

In the sections above, we initially selected 8 different days (a three day run, and five individual days) to demonstrate the performance of the revised CACTus algorithm. Using this version of CACTus, we have also generated a full CME catalog, extending from 2007 to 2014, which is available at \url{http://www.sidc.be/cactus/hi} and \url{http://www.helcats-fp7.eu/catalogues/wp2_cactus.html}. The methodology employed to generate the full catalog (i.e, the thresholds in intensity, etc) is identical to what has been described above, except that the analysis is performed on intervals of one month duration, which overlap by one day in order to accommodate CMEs that span the month boundary. A limit of one month is imposed for computational reasons. It is important to note that the length of time interval that we incorporate into a single CACTus analysis run will affect the detection and/or returned parameters of a CME since the intensity threshold in accumulator  space ($W_{thresh}$) depends on $W_{mean}$ and $W_{sd}$, both of which depend on the duration of the analysed interval. We choose 15 different ``good" and ``fair" quality CMEs from the HELCATS manual catalog, from different times between 2007 and 2013 and with a range of values of $\phi_{SSE}$ (see table~\ref{table4}). Table~\ref{table4} also includes the events listed in table~\ref{table2} (shown in boldface).
\begin{itemize}
\item \textbf{{\it Comparison with CMEs discussed in section \ref{mancatalog}}}: To demonstrate the effect of the length of the analysed interval, we compare the results for the six CMEs listed in table~\ref{table2} with their entries in the online CACTus catalog. The dates and times of these CMEs are highlighted in boldface in table~\ref{table4}. We notice that there are, obviously, some differences that arise when an interval of one month is incorporated in a single analysis. 
\begin{itemize}
\item Of the six CMEs listed in table~\ref{table2}, the times of appearance of four match well with their counterparts listed in table~\ref{table4} (in boldface). For other two CMEs, there is a difference of 2 frames (corresponding to 80 minutes). As discussed above, this may be attributed to the different thresholds resulting from the analysis of different lengths of data.
\item Out of the six CMEs listed in table~\ref{table2}, the PA widths of the CMEs detected on 2010/04/03 and 2007/04/19 match well with those listed in table~\ref{table4}.  For the CMEs detected on 2008/02/04 and 2007/10/22, the PA width quoted in table~\ref{table4} is less than the PA width quoted in table~\ref{table2}. On closer inspection we find that, when using a longer analysis interval, each of these CMEs was identified as two separate (narrower) CMEs. 
Reanalysis results in the CME on 2008/02/04 being identified as two separate CMEs because the leading edge of the CME is poorly defined. We suggest that, in this case, the slight change in the intensity threshold due to the longer analysis interval results in the points in accumulator space not being joined up under the application of the morphological closing technique. Hence the algorithm detects two narrower CMEs instead of a single wider one. Finally, for the CMEs detected on 2008/12/12 and 2008/04/26, the PA widths of the CMEs in table~\ref{table4} are greater than their counterparts in table~\ref{table2}. This can also be explained in terms of there being a different threshold due to increased length of the analysed dataset, which results in the inclusion of additional material (possibly the CME flanks or adjacent structures). For these two CMEs, the PA widths quoted in the online CACTus catalog are actually much more consistent with those in the manual catalog. 
\item Different lengths of analysis intervals will clearly also result in different CME velocities. CACTus computes the maximum velocity in each position angle bin within a CME. The velocity listed in the online CACTus catalog is the median of all these maximum velocities. The median is likely to change if the PA width of the CME changes. Moreover, a different threshold is likely to give rise to a different estimate of the maximum velocity at a given PA (see section~\ref{estimateparam}). It is likely that both of these effects would contribute to differences in the CME velocities if the length of the analysis interval is modified significantly. However we find that the differences in the velocities of the common CMEs in table~\ref{table2} and table~\ref{table4} are within one standard deviation (dv, quoted in table~\ref{table1}).
\end{itemize}

It is difficult to ascertain whether it is better to analyse individual days (or short intervals) or longer periods of data (such as a month). However, it is important to employ a consistent methodology. Therefore, for in the generation of the online catalog, we incorporate one month of data into each CACTus run from 2007 to 2014.
\item \textbf{{\it Comparison with manual catalog}}: Here we compare the online CACTus catalog with the manual catalog. We compare every event listed in both manual and automated catalogs on the selected days. A detailed comparison between the manual and the automated catalogs will be presented in future studies.

\begin{itemize}
\item The times of CME appearance in the two catalogs differ at most by 3 frames (2 hours).  In section~\ref{timecomparison}, we have compared the manual and automatically-derived parameters of the CME that entered the FOV of HI-1 on STEREO-A on 2007/04/19. The times at which the CME on 2008/04/26 was first detected by CACTus is two hours earlier than its manual catalog entry. We suggest that this discrepancy is due to the outflow of associated material prior to the CME itself (see animated figure~\ref{fig9} available online). 
Sometimes the leading edge of a CME is faint or not well defined, for example the CME on 2011/02/26, which makes it difficult to estimate accurately its time of first appearance in the FOV (see animated figure~\ref{fig10} available online). For some CMEs, the time of appearance match very well in both the automated and manual catalogs. Such events tend to have bright clear leading edges (for example the CME on 2013/02/27) and/or are preceded by outflows that are faint enough not to be detected by CACTus (see animated figure~\ref{fig7} available online). In general, we suggest that CACTus tends to detect CMEs either at the same time or slightly earlier than is estimated by manual operators.
\item In general, CACTus tends to produce a lower estimate of the PA width than is included in the manual catalog for the same event. As noted previously, this can be due to the mis-detection by CACTus of a single CME as two narrower events or can be due to the fact that CMEs flanks can be faint and can sometimes move non radially.
\item Velocities estimated by CACTus tend to match well with $v_{FP}$ in the HELCATS manual catalog for limb/near-limb CMEs (those with $\phi_{SSE}$ close to 90$\degree$).
\item For the days considered in table~\ref{table4}, we find that CACTus detects a total of 18 events. The manual catalog includes only 15 events, meaning that CACTus detects 3 events that are not listed in the manual catalog. One of these additional events (on 2013/02/27) is due to CACTus identifying, as a separate CME, pre-CME outflows. Another is due to the erroneous splitting of a single CME into two (2007/10/22). Finally one narrow CME is also identified that is not listed in the manual catalog as it is narrower than the PA width threshold of 20$\degree$ imposed in the generation of that catalog.
\end{itemize} 

\item {\it Corotating Interaction Regions (CIRs):} CIRs are the regions of high density plasma that form at the interface between the fast and the slow solar wind  by compression \citep{1999SSRv...89...21G}. \citet{2008GeoRL..3510110R} and \citet{2008ApJ...675..853S,2008ApJ...674L.109S} first demonstrated white-light imaging of CIRs in the heliosphere using STEREO/HI. However, the authors suggested that HI was actually observing pre-existing streamer blobs that had become entrained at the stream interface. Compression of the blobs at the stream interface means that they can be clearly tracked out to 1 AU and beyond. Although it is thought that this entrainment happens beyond the HI-1 FOV \citep[see][]{2016SoPh..291.1853P}, we have examined the CACTus output to see if it has detected any of the CIR-associated blobs presented in the literature. We find that CACTus detects some, but not all, of these features. For example CACTus detects two of the six CIR-associated blobs that were tracked by \citet{2010JGRA..115.4103R} between 2007/09/09 and 2007/09/12, but none of the six CIR-associated blobs tracked by those authors between 2007/09/17 and 2007/09/20. Similarly, \citet{2009ApJ...702..862T} detected the tracks of several CIR-associated blobs between 2008/11/14 and  2011/11/17, only one of which was detected by CACTus. The CIR-entrained blobs detected by CACTus have PA widths that are less than 20$\degree$. Therefore, a PA threshold of 20$\degree$ would likely exclude many such blobs. However, it may also remove some features that are associated with CMEs. The detection of CIR-associated blobs using CACTus will be the subject of a future study.
\end{itemize}
\begin{table}
\begin{center}
\caption{Comparison of online CACTus catalog with manually generated catalog}  

\label{table4}
\begin{tabular}{c l l l | c l l l l l l l}
\hline \hline
\multicolumn{4}{c}{CACTus Online} & \multicolumn{8}{c}{Manual} \\
t$_{0}$ & CPA & da & v  &  t$_{0}$ & CPA & da & v$_{SSE}$ & $\phi_{SSE}$ & v$_{FP}$ & v$_{HM}$ & type \\
\hline
2013/05/13 20:09 & 87 & 94 & 745 & 2013/05/13 20:09 & 87 & 135 & 738 & 81 & 705 & 769 & good\\
\hline
2013/02/27 15:29 & 83 & 82 & 411 & Extra event identified by CACTus. & &  &  & &  & & \\
2013/02/27 10:09 & 96 & 88 & 520 & 2013/02/27 10:09 & 95 & 130 & 597 & 35 & 577 & 612 & fair\\
\hline
2012/07/02 21:29 & 115 & 42 & 502 & 2012/07/02 20:49 & 100 & 100 & 434 & 75 & 426 & 441 & good\\
2012/02/06 02:49 & 67 & 46 & 414 & 2012/02/06 02:49 & 63 & 65 & 410 & 95 & 399 & 420 & fair\\
2011/06/01 22:09 & 88 & 60 & 505 & 2011/06/01 22:49 & 92 & 85 & 774 & 139 & 667 & 931 & good\\
2011/02/26 04:09 & 82 & 64 & 352 & 2011/02/26 05:29 & 72 & 85 & 389 & 48 & 383 & 392 & fair\\
2010/06/03 11:29 & 88 & 64 & 307 & 2010/06/03 12:09 & 90 & 120 & 333 & 100 & 320 & 347 & good\\
\textbf{2010/04/03 12:09} & 102  & 72 & 823 & 2010/04/03 12:09 & 102 & 105 & 927 & 77 & 889 &962 & good \\
2009/07/31 12:49 & 82 & 36 & 334 & 2009/07/31 13:29 & 40 & 40 & 331 & 70 & 326 & 334 & fair \\
2009/01/09 07:29 & 92 & 48 & 330 & 2009/01/09 08:49 & 88 & 65 & 522 & 26 & 480 & 570 & good\\
\textbf{2008/12/12 14:09}& 77& 78& 475& 2008/12/12 15:29 & 77 & 95 & 426 & 69 & 419 & 431 & good\\
\textbf{2008/04/26 16:49}& 91& 98& 553& 2008/04/26 18:49 & 85 & 110 & 649 & 95 & 620 & 679 & good\\
\hline

\textbf{2008/02/04 14:09}& 96& 40& 437& 2008/02/04 14:09 & 77 & 115 & 511 & 53 & 506 & 515 & good\\
2008/02/04 03:29& 75& 30& 505&  Not listed in catalog& & & & & &\\
\hline
2007/10/22 19:29& 74& 24& 224& Single event is detected as && & &&& \\
& & & & two different events by CACTus & & & &&& \\
\textbf{2007/10/22 14:49}& 92& 20& 218& 2007/10/22 14:49 & 77 & 85 & 691 & 174 & 499 & 748 & fair\\
\hline
\textbf{2007/04/19 11:30}& 91& 70& 371& 2007/04/19 13:30 & 90 & 100 & 392 & 61 & 389 & 393 & good\\
\hline

\end{tabular}
\end{center}
\end{table}

\section{Summary and Discussion}\label{SECT:Discussion}
We have applied, with success, the CACTus algorithm, to automatically detect CMEs in images from the HI-1 camera on-board STEREO-A. We performed the analysis on 1 day background-subtracted level 2 data. We have experimented with different background-subtraction methodologies and find it makes little difference because, ultimately, we use polar transformed running difference images for detection using the Hough transform. In summary, the different background subtraction techniques do not significantly affect the performance of CACTus especially for bright CMEs. 

We have also applied this algorithm to HI-1 images from STEREO-B but with less success. The HI-1 camera on STEREO-B suffers from pointing anomalies due, it is now thought, to a small amount of mechanical instability in the attachment of the camera to the HI instrument structure \citep{2009SoPh..254..185B,2012MNRAS.420.1355D,2016SoPh}. Given the large gradients inherent in the F-corona, which has a much larger white-light signal than that of CMEs, such pointing jitter can result in the presence of a residual F-coronal signal in the STEREO-B HI-1 imagery. This residual signal detrimentally impacts the automatic detection, in particular, of CMEs by CACTus. Hence, more advanced techniques are required to detect CMEs reliably in HI-1 images from STEREO-B. Here we primarily present the technical details of the detection algorithm that we implemented within CACTus. 
Of the 8 days initially studied here, 3 days (2010/04/02 to 2010/04/04) were analysed as a single run and the remaining 5 days were analysed individually. 
Incorporating these 5 individually-analysed days (see table~\ref{table2}) into monthly runs (see table~\ref{table4}) makes little difference to the CACTus results. Any differences are due to the different thresholds that result from the incorporation of one month data.
We also compare a few additional events taken at random from the online CACTus catalog (these encompass both ``good" and ``fair" CMEs as defined by the HELCATS manual catalogers). We compared the automatically and manually-derived CME properties (see, table~\ref{table4}).

We find that, at least over the limited number of days presented in the current paper, CACTus detects more events in HI-1 images than are detected by the manual cataloging. This is due to several factors that are inherent in the method (the selection of the intensity threshold value and kernel size for morphological closing). In this study some of the events detected by CACTus are excluded by human operators due to different thresholds in PA width extent. Manual cataloging endeavours need to apply a relatively large PA width threshold that is not necessary with an automated detection routine such as CACTus, simply due to the arduous nature of the manual-cataloging process. It is worth noting that even when we apply objective definitions, there are instrumental factors that come into play even in automatic CME detection. For example, the apparent brightness of a CME in white-light imagery is dependent to some degree on its location relative to the Thomson sphere. Also the application of a PA width threshold in order to identify a CME is not entirely consistent as PA width is a projection of a CME's true width onto the POS. Conversely, there are some benefits to a manual as opposed to an entirely algorithmic approach. For example, what is clearly a single CME in the HI imagery on 2007/10/22 (and in the manual catalog) is detected as multiple (two separate) events by CACTus due to the large variation in the brightness over the CME front. In fact, angularly extended faint CME fronts (as characterised in the manual catalog) commonly appear as several narrower CMEs in the CACTus catalog. This tendency for CACTus to overestimate the number of CMEs is also seen in the LASCO and COR2 implementations of CACTus, so it is an intrinsic problem of the method \citep{2009ApJ...691.1222R}.

We also note that CACTus sometimes computes the time of appearance of a CME to be earlier than it is in the manual catalog. This can be explained by the presence of pre-CME outflows or the lack of well defined and bright leading edge. We feel that the differences in estimation of the time of first appearance of a CME do not necessarily mean that an automated algorithm is superior to the discretion of manual operators. Rather, it indicates the lack of a proper definition of the time of appearance of a CME.

We find that the PA width of a given CME estimated by CACTus tends to be lower than is listed in the manual catalog. This effect can be exacerbated by the presence of a faint shock around the CME that is often taken into account by manual catalogers but excluded most of the time CACTus. This discrepancy in CME width was also present in earlier version of CACTus \citep[see][]{2009ApJ...691.1222R}. The difference in width measurement is also due to the lack of a proper definition of the width of a CME. \citet{2004A&A...422..307C} reported a difference of a few degrees to more than a hundred degrees when comparing manual catalogs with their measurements. It is still subjective whether, or not, shocks should be included when estimating the width of a CME. CACTus implements a somewhat more objective criteria for this but, still, the shocks around the flanks of CMEs may or may not be included. The time of CME appearance from CACTus is, on average, earlier than that quoted in the manual catalog. A comprehensive comparison with the manual catalog is underway and will be presented in a future paper.\\
Currently CACTus yields only the velocity projected onto the POS as seen from the observing spacecraft. This should be borne in mind when using the speeds derived by CACTus for specific research purposes. Options for retrieving the 3D speed from CACTus may be sought in the future, for  example through the inclusion of a more realistic estimate of the propagation direction (such as from GCS or the CAT tools). However, this may pose potential problems, in particular for its real-time implementation. It may also be that CACTus could be optimised to extract the curved ridges in time-elongation maps from heliospheric imaging observations that extend out to large elongation; these tracks contain information regarding the 3D propagation direction within their curvature.\\
One distinct advantage of automated detection is that it is not affected by the vagaries of human subjectivity and, hence, the detection of CMEs will be more consistent. As noted above, one disadvantage of this is that it generally results in the detection of many more events than are listed in manually compiled catalog. However, the number of detected events depends critically on the threshold value of brightness implemented in the program. Optimisation of this value of this threshold for HI imagery will be evaluated in the future versions of CACTus for HI. \\
To summarise, in this work we have described the basic detection algorithm for detecting CMEs in heliospheric imager data using CACTus for the first time. This is of great importance for heliospheric physics. To best of our knowledge, this is the first report on the successful fully-automated detection of CMES in data from the STEREO heliospheric imagers HI-1.

\section{Acknowledgments}
The authors thank the referee for his/her valuable in-depth comments which have helped us to improve the manuscript. V.P. acknowledges the support of the entire CACTus team at the Royal Observatory of Belgium. This project has received funding from the European Union's Seventh Framework Programme for research, technological development and demonstration under grant agreement no 606692 [HELCATS]. We acknowledge support from the Belgian Federal Science Policy Office (BELSPO) through the ESA-PRODEX program and from the Interuniversity Attraction Poles Programme initiated by the Belgian Science Policy Office (IAP P7/08 CHARM).


\end{document}